\documentclass{article}

\usepackage{PRIMEarxiv}

\usepackage[utf8]{inputenc} 
\usepackage[T1]{fontenc}    
\usepackage{hyperref}       
\usepackage{url}            
\usepackage{booktabs}       
\usepackage{amsfonts}       
\usepackage{nicefrac}       
\usepackage{microtype}      
\usepackage{lipsum}
\usepackage{fancyhdr}       
\usepackage{graphicx}       
\graphicspath{{media/}}     
\usepackage{csquotes} 

\usepackage[utf8]{inputenc}
\usepackage{amssymb,amsmath,amsthm} 
\usepackage[scr=boondox]{mathalfa} 
\usepackage[french]{babel}
\usepackage{xspace}

\newcommand{\nospacecolon}{\unskip:\ignorespaces}

\usepackage[utf8]{inputenc}
\usepackage{lmodern}

\usepackage{tabularx} 
\usepackage{graphicx}
\usepackage{adjustbox}
\usepackage{longtable}
\usepackage{array}
\usepackage{ragged2e}
\usepackage{caption}
\usepackage{multirow}
\usepackage{blindtext}
\usepackage{multicol}

\usepackage{float} 
\usepackage[usenames,dvipsnames]{xcolor} 
\usepackage{enumitem} 
\usepackage{pgfplots} 
\pgfplotsset{compat=newest}
\pgfplotsset{width=7.5cm}
\pgfkeys{/pgf/number format/1000 sep={}}
\usepackage{tikz} 
\usetikzlibrary{shapes.geometric,calc, positioning,arrows}
\usetikzlibrary{decorations.pathreplacing}
\usetikzlibrary{fadings}
\usepackage{subcaption}
\usepackage{makecell}
\usepackage{multirow}
\usepackage{rotating}
\usepackage{diagbox}
\usepackage{wrapfig}

\addto\captionsfrench{}

\newcommand{\EUT}[1]{\mathbb{E}[u(#1)]}
\newcommand{\R}{\mathbb{R}}
\newcommand{\ou}{\mathscr{o}}
\newcommand{\dkl}[2]{D_{\text{KL}}(#1 || #2)}

\newcommand{\suchthat}{\;\ifnum\currentgrouptype=16 \middle\fi|\;}
\newcommand\footnoteref[1]{\protected@xdef\@thefnmark{\ref{#1}}\@footnotemark}
\DeclareMathOperator*{\maxi}{\text{máx}}

\DeclareMathOperator*{\argmax}{\text{argmax}}
\DeclareMathOperator*{\argmin}{\text{argmin}}
\DeclareFontFamily{U}{mathx}{\hyphenchar\font45}
\DeclareFontShape{U}{mathx}{m}{n}{
      <5> <6> <7> <8> <9> <10>
      <10.95> <12> <14.4> <17.28> <20.74> <24.88>
      mathx10
      }{}
\DeclareSymbolFont{mathx}{U}{mathx}{m}{n}
\DeclareMathSymbol{\bigtimes}{1}{mathx}{"91}
\title{Sports Betting: an application of neural networks and modern portfolio theory to the English Premier League

}
\author{
  Román Alberto Vélez Jiménez \\
  Instituto Tecnológico Autónomo de México \\
  Mexico City\\
  \texttt{roman.velez@itam.mx} \\
   \And
  José Manuel Lecuanda Ontiveros \\
  Centro de Enseñanza Técnica y Superior  \\
  Mexico City\\
  \texttt{manuel.lecuanda@cetys.mx} \\
   \And
  Edgar Possani \\
  Instituto Tecnológico Autónomo de México \\
  Mexico City\\
  \texttt{epossani@itam.mx} \\
}

\begin{document}
\maketitle 

\begin{abstract}
This paper presents a novel approach for optimizing betting strategies in sports gambling by integrating Von Neumann-Morgenstern Expected Utility Theory, deep learning techniques, and advanced formulations of the Kelly Criterion. By combining neural network models with portfolio optimization, our method achieved remarkable profits of 135.8\% relative to the initial wealth during the latter half of the 20/21 season of the English Premier League. We explore complete and restricted strategies, evaluating their performance, risk management, and diversification. A deep neural network model is developed to forecast match outcomes, addressing challenges such as limited variables. Our research provides valuable insights and practical applications in the field of sports betting and predictive modeling.

\end{abstract}
\section{Introduction}

This study aims to address the question of which bets a rational gambler, considering their risk appetite, should select in order to maximize expected utility. The concept of optimality is defined based on the Von Neumann-Morgenstern Classical Utility Theory. We employ deep learning techniques to estimate event odds, and subsequently employ the Sharpe Ratio and Kelly's financial criteria to identify the optimal set of bets.

The primary objective of this study is to assess the performance of the Sharpe Ratio and the Kelly Criterion within the context of real-world sports betting, specifically during the second half of the 2020-2021 season of the English Premier League (EPL). Additionally, compare the impact of underlying assumptions on the aforementioned criteria, shedding light on their significance.

\subsection{Sports Betting}
The analysis of a bet can be approached as a decision problem defined by the tuple $(D, E, C, (\succeq))$. Within our mathematical framework, this decision tuple $(D, E, C, (\succeq))$ encompasses fundamental elements for making rational decisions within the realm of sports betting. The decision set $D$ represents the available choices or bets that a bettor can select. The events set $E$ defines the potential outcomes or events associated with these bets. The consequence set $C$ encompasses the various outcomes or consequences that arise from each event, denoted as $c = c(d, e)$. The preference relation $(\succeq)$ captures the bettor's subjective preferences and enables comparisons between different bets based on personal criteria \cite{de2017theory}. By defining and analyzing these elements, our approach aims to assist rational gamblers in choosing optimal bets that maximize their expected utility and align with their preferences \cite{mendoza2011estadistica}. Specifically, we seek the optimal bet $d_{*} \in D$ based on the bettor's preferences $(\succeq)$. These preferences are contingent upon the consequences $c \in C$, which in this case include the odds and probability of the event $e \in E$ \cite{zimmermann2013predicting}. 

We denote $\ell$ as the percentage of the wealth wagered on an event $e$, hence $\underline{\ell}$ represents the wager vector. Consequently, the decision $d_{\underline{\ell}}$ signifies the act of placing a bet of $\underline{\ell}$ on the events $e \in E$.

\subsubsection{Odds}
Within the domain of sports betting, let $W_{0}$ represent the initial wealth and $W_{1}$ denote the wealth obtained for placing a bet on event $e$ but not taking in account the initial quantity $W_{0}$. In the event's occurrence, the outcome manifests as a gross profit of $W_{1} + W_{0}$, while in its non-occurrence, the bet is forfeited, resulting in a loss. The probabilities associated with these outcomes are denoted by $p$ and $1-p$ respectively. It is assumed that all wagers result in a positive return, that is, $W_{1} > 0$.

On the other hand, the odds $\ou$ are defined as the ratio between the gross profit and the initial wealth, i.e. $\ou = (W_{1} + W_{0})/W_{0} = 1 + W_{1}/W_{0}$. It is \textit{assumed that the odds and probabilities of the events are fixed over time}.

\subsubsection{Returns}
Formally, the net return $\varrho$ from betting \$1 on event $e$ is the random variable $\varrho = \ou - 1$ with probability $p$ and $\varrho = - 1$ with probability $1-p$. In the case where there are $m$ events, it is defined as the random vector $\underline{\varrho}$, where the $i$-th entry is the random variable $\varrho_{i}$. 
To facilitate notation, we introduce the odds vector $\underline{\ou}:= (\ou_{1}, \ou_{2}, \dots, \ou_{m})'$, representing the individual odds per event. Similarly, we define the probability vector $\underline{p}$ in the same manner. Additionally, we denote the diagonal matrix of odds as $D_{\ou} := \text{diag}(\ou_{1}, \ou_{2}, \dots, \ou_{m})$.
\begin{equation} \label{eq:1.ganancia_neta}
    \text{i.e.} \quad \underline{\varrho} = D_{\ou} \mathscr{m} - \mathbf{1}, \quad \mathscr{m} \sim \text{multinomial}(1; \underline{p}).
\end{equation}

Likewise, the total return $R(\underline{\ell})$ of one hundred percent of the initial wealth for betting $\ell_{i}$ percent of the wealth to the event $e_{i}$ is equal to the initial wealth plus the random gains or consequences \cite{whitrow2007algorithms}. In mathematical notation,
\begin{equation} \label{eq:1.ganancia_total}
    R(\underline{\ell}) = 1 + \sum_{i=1}^{m}\ell_{i} \varrho_{i} = 1 + \underline{\ell}' \underline{\varrho}.
\end{equation}

The analysis assumes certain conditions regarding the nature of the betting system. Specifically, it assumes the absence of \textit{short selling} and borrowing, and assumes that money is infinitely divisible with no minimum bet requirement. These assumptions can be summarized as follows: First, it is implied that the wager amount $\ell_{i}$ for each event $i$ is non-negative ($\ell_{i} \geq 0$) for all $i$. Second, the assumption is made that the total wager across all events satisfies the constraint $\sum_{i}^{m} \ell_{i} \leq 1$. Lastly, the last two assumptions state that the wager amount $\ell_{i}$ for each event $i$ is bounded within the range $[0,1]$.

\subsubsection{Betting Market} \label{sct:1.1MercadoApuestas}
By definition, for a bet to be fair one would expect the net return to be zero, i.e. $\mathbb{E}[\varrho] = 0$. The above happens if and only if $\ou = 1/p$. But, the odds of a bookmaker are always of the form $\sum_{i}^{m}1/\ou_{i} = 1 + tt$ \cite{sumpter2016soccermatics}, where the $tt$ is known as the commission (commonly known as track take) that a casino charges. Therefore $tt > 0$. However, since the betting market is a non-efficient market \cite{jakobsson2007testing}, one can find odds $\underline{\ou}$ of different casinos such that $tt \leq 0$. When the market commission is negative, arbitrage is obtained \footnote{By arbitrage, it is mean that there is a price advantage between bookmakers such that, regardless of the outcome, a fixed strategy \textit{always} makes money.} . The strategy that exploits this phenomenon is of the form $\ell_{i}^{(A)} := \ou_{i}^{-1}/\sum_{j}\ou_{j}^{-1}$, since fixing the event $e_{i}$ the total return $R(\underline{\ell}_{A}) = 1 + \ell_{i}^{(A)} \ou_{i} - \sum_{j}^{m} \ell_{j}^{(A)} = 1/(1+tt), \quad \forall e_{i} \in E$. It will be shown below, that, this strategy matches the vector of strategies found by the Sharpe Ratio Criterion for non-efficient markets.


\subsection{Uncertainty}
This section provides a compilation of key findings from Frequentist Statistical Inference and Information Theory, which serve as justifications for the methods employed in this research.
\subsubsection{Information Theory}
In information theory, the concept of entropy is introduced to quantify the uncertainty associated with a phenomenon represented by the random variable $X$. Entropy, denoted as $H(X)$, is defined as $H(X) := - \mathbb{E}_{F}[\log(X)] = - \int{\mathcal{X}}\log d F(x)$, where the random variable $X$ follows a distribution characterized by $F$ \cite{shannon2001mathematical}. In other words, we express $X$ as $X \sim F$. It is worth noting that the density function $f$ is related to the distribution function $F$ through the expression $f(x) = \frac{d}{dx}F(x)$. The entropy tends to approach $0^+$ when uncertainty is low and approaches infinity in the presence of high uncertainty. Cross-entropy, denoted as $H(F, G) := - \mathbb{E}_{F}[\log(g(X))]$, quantifies the difference between distributions $F$ and $G$ for the same phenomenon. Furthermore, the Kullback-Leibler Divergence (KL-Divergence) \cite{kullback1997information} provides a means to estimate the disparity between distributions $F$ and $G$, which is given by 
\begin{equation}\label{eq:1kld}
    \dkl{F}{G} := \mathbb{E}_{F}\left[ \log \left(f(X)/g(X) \right) \right].
\end{equation}
Two important properties of the KL-Divergence are its direct relationship to cross-entropy and its equivalence to the minimization of KL-Divergence when maximizing the likelihood of a random sample. This equivalence is demonstrated through $\max\{L(\underline{\theta}; \underline{x})\} = 
\max \left \{\frac{1}{n}\sum_{j}^{n}\log(f(\underline{x}_{j}; \underline{\theta}))  \right\} = \max \left \{ \mathbb{E}_{\hat{F}_{\text{emp}}}[\log(f(\underline{x}_{j}; \underline{\theta}))] \right\} = \min \{H(\hat{F}_{\text{emp}}, F)\} = \min \{\dkl{\hat{F}_{\text{emp}}}{F}\}$ \cite{nielsen2015neural}.

\subsubsection{Deep Learning}
In the realm of event prediction, deep learning methods have gained prominence due to their effectiveness in nonlinear scenarios. They excel in identifying relationships between covariates and exhibit flexibility in optimizing objective functions to address diverse requirements for the same phenomenon \cite{nielsen2015neural}. For instance, optimizing the KL-Divergence for observed and predicted data is a valuable tool employed to estimate the odds of EPL matches in the study.

With this approach, it becomes possible to make more informed wagers with bookmakers. However, to complete the process, it is essential to determine the specific events to bet on and the corresponding wager allocation based not only on the probabilities estimated but also on the market odds.


\subsection{Utility Theory}\label{sct1:utility}
In order to avoid contradictions and paradoxes  associated with the moral value given to money, this study adopts the Von Neumann-Morgenstern Utility Axioms \cite{von2007theory} as the basis for its methodology. Leveraging this theory offers several notable advantages, including the recognition that the nominal value of money differs from its moral value, as well as the establishment of a clear correspondence between qualitative preferences and quantitative utilities in the context of gambling.

Consider a probability space $(\Omega, \mathscr{F}, \mathbb{P})$, where $\Omega$ represents the sample space, $\mathscr{F}$ denotes the associated sigma algebra, and $\mathbb{P}$ is the probability measure. Within this context, let $X$ be a random variable that follows a distribution characterized by $F$, and takes on values $x$ belonging to the set $\mathcal{X}$. The probability distribution $F$ is referred to as a "lottery" over the set $\mathcal{X}$. The objective is to establish preference relations $(\succeq)$ over the set of lotteries, denoted by $\mathscr{L} = \{F | F \text{ is a probability distribution over } \mathcal{X}\}$ \cite{von2007theory}. In essence, making a decision $d \in D$ to choose a lottery $F \in \mathscr{L}$ is tantamount to studying $D$ itself, given the fixed probabilities. Relaxing the assumption of fixed probabilities would transform the gambling problem into a Bayesian framework \cite{mendoza2011estadistica}.

As previously mentioned, the moral value of money differs from its nominal value \cite{kahneman2003perspective} \cite{bernoulli1954exposition}. To capture this distinction, the moral value associated with a monetary outcome $x$ is modeled using a Bernoulli Utility Function $u\nospacecolon\mathcal{X} \rightarrow \mathbb{R}$. Similarly, to quantify the utility of a lottery $F$, a Von Neumann-Morgenstern Utility Functional $U\nospacecolon\mathcal{X} \rightarrow \mathbb{R}$ is employed. According to the Expected Utility Theorem \cite{von2007theory}, the utility functional can be expressed as $U(F) = \int_{\mathcal{X}} u dF = \mathbb{E}_{F}[u(X)]$. Additionally, it follows that $U(F_X) \geq U(F_Y) \iff F_X \succeq F_Y$ \cite{von2007theory}. Consequently, a gambler's preferences can be quantified through the utility function $u$, which is determined by the individual's risk profile.

\subsubsection{Utility Functions}
In economic practice, the utility function $u$ is commonly assumed to be an increasing function that exhibits decreasing marginal rates of substitution. This implies that $u'(x) > 0$ and $u''(x) < 0$ for $x \in \mathcal{X}$. Such assumptions capture the notion that individuals with higher wealth tend to exhibit lower levels of risk aversion \cite{kahneman2011thinking}.

Modern Portfolio Theory, pioneered by Harry Markowitz, suppose a quadratic utility function, where $u(x)$ is a polynomial of degree two, given by $u(x) = \beta_0 + \beta_1 x + \beta_2 x^2$ \cite{markowitz1968portfolio}. Consequently, the utility of a lottery can be expressed as \nospacecolon
\begin{equation} \label{eq:1.quadratic_utility}
U(F) = \mathbb{E}_F[X] = \beta_0 + \beta_1 \mathbb{E}[X] + \beta_2 (\text{Var}(X) + \mathbb{E}[X]^2) = W(\mu, \sigma).
\end{equation}
Here, $W$ denotes a utility function that depends on the mean $\mu$ and variance $\sigma^2$ of the random variable $X$. Thus, the utility of a lottery can be fully characterized by its mean and variance. In fact, $F_1 \succeq F_2 \iff W(\mu_1, \sigma_1) \geq W(\mu_2, \sigma_2)$. To align with the principle that greater wealth is always preferred, the function $W$ must be monotonically increasing in $\mu$, while also being monotonically decreasing in $\sigma$. In other words, for the same expected return, individuals with risk-averse preferences prefer lotteries with lower variance.

On another note, Daniel Bernoulli argued in his work "Exposition of a New Theory of the Measurement of Risk" that the change in utility experienced by an individual is inversely proportional to their wealth \cite{bernoulli1954exposition}. Consequently, Bernoulli suggested that utility functions follow a logarithmic form\nospacecolon
\begin{equation}\label{eq:1.log_utility}
u'(x) = \frac{1}{x} \implies u(x) = \log(x) + C.
\end{equation}
Once the utility functions have been characterized and the probabilities of the events have been estimated, the objective is to identify the lottery that maximizes expected utility in order to determine the optimal bet.
\section{Betting Strategies}

In this section, we address the problem of determining the best strategy for a set of $r$ random rewards within the betting system described previously.

\subsection{Sharpe Ratio}
Under the assumption (\ref{eq:1.quadratic_utility}) that a rational gambler's utility follows a quadratic form, the objective is to identify the best strategy in the universe $\Psi = \{\psi = (\mu, \sigma)| \sum_{k = 1}^{r} \ell_{k} = 1\}$ \cite{markowitz1968portfolio}. Here, $\mu = \sum_{k}^{r} \ell_{k} \mu_k$ represents the return of the portfolio, and $\sigma^{2} = \underline{\ell}' \Sigma \underline{\ell}$ denotes the portfolio's variance. The covariance matrix $\Sigma$ captures the covariances between the $r$ random rewards, while $\mu_{k} = \mathbb{E}[\varrho_{k}]$ represents the expected value of bet $k$.

A rational strategy aims to minimize the portfolio variance while maintaining a specified expected return level $\mu_{*}$. Such a strategy $\underline{\ell}_{*}$ can be obtained by solving the following optimization problem\nospacecolon
\begin{equation} \label{eq:3.min_fixedreturn}
    \argmin_{\substack{\underline{\ell} \geq \mathbf{0}}} \{ \underline{\ell}'  \Sigma \underline{\ell}\} \quad \text{subject to} \quad 
        \underline{\ell}' \underline{\mu} = \mu_{*}, \quad
        \sum_{k}^{r} \ell_k = 1.
\end{equation}

In the context of sports gambling, where simultaneous bets are placed, the portfolio return is given by $\underline{\mu} = D_{\underline{\ou}}\underline{p} - \mathbf{1}$, and the covariance matrix is $\Sigma = D_{\ou}\left(\text{diag}(\underline{p}) - \underline{p} \underline{p}'\right) D_{\ou}$, as shown in equation (\ref{eq:1.ganancia_neta}). In the case of simultaneous bets, $\Sigma$ becomes a block diagonal matrix, denoted as $\Sigma = \text{diag}\left(\Sigma_{1}, \Sigma_{2}, \dots, \Sigma_{r}\right)$.

The set of optimal portfolios with minimum variance, for all possible return levels, is referred to as the Efficient Frontier \cite{elbannan2015capital}. To identify the best portfolio within this optimal set, the Sharpe Ratio was used, which is defined as the ratio between the difference of the portfolio return and the risk-free rate $R_{f}$, and the standard deviation of the portfolio \cite{sharpe1998sharpe}. The Sharpe Ratio is given by\nospacecolon
$
S(\underline{\ell}) := (\underline{\ell}' \underline{\mu} - R_{f})/\sqrt{\underline{\ell}' \Sigma \underline{\ell}}.
$
To facilitate the optimization process, it is beneficial to transform the Sharpe Ratio problem into a convex optimization problem by introducing an additional dimension. This transformation helps avoid issues related to non-convexity and local optima \cite{cornuejols2006optimization}. It is introduced a change of variable $\underline{y} = \kappa \underline{\ell}$, assuming a feasible solution exists such that $\underline{\ell}' \underline{\mu} > R_{f}$, and fix a scalar $\kappa > 0$ such that $(\underline{\mu} - R_{f}\mathbf{1})'\underline{y} = 1$. The resulting convex optimization problem in $r+1$ dimensions is as follows \cite{cornuejols2006optimization}\nospacecolon

\begin{equation} \label{eq:2.min_sharperatio_convex}
\argmin_{\substack{\underline{y} \geq \mathbf{0}}} \left \{ \underline{y}' \Sigma \underline{y} \right \} \quad \text{subject to} \quad 
    \begin{matrix}
        \left(\underline{\mu} - R_{f}\mathbf{1}\right)'\underline{y} = 1 \\
        \quad \sum_{k}^{r}y_{k} = \kappa \\
        \kappa > 0
    \end{matrix} .
\end{equation}

As mentioned in Section \ref{sct:1.1MercadoApuestas}, in a sports betting market where the commission $tt$ is negative and under the strategy $\underline{\ell}_{A}$, the gross return $R(\underline{\ell}_{A}) > 1$. The Sharpe Criterion helps identify such market inconsistencies by converging on the optimal strategy $\ell_{A}$. This occurs because $\mathbb{E}[R(\underline{\ell}_{A})] = \mathbb{E}[1 + \underline{\ell}_{A}' \underline{\varrho}] = 1 + \underline{\ell}_{A}'(D{\ou}\underline{p} - \mathbf{1}) = 1/(1 + tt) > 1$. Furthermore, $\text{Var}(R(\underline{\ell}_{A})) = \underline{\ell}_{A}' \text{Var}(D_{\ou} \mathscr{m} - \mathbf{1}) \underline{\ell}_{A} = \underline{\ell}_{A}' (D_{\ou}(\text{diag}(\underline{p}) - \underline{p} \underline{p}') D_{\ou}) \underline{\ell}_{A} = 0$. Thus, the optimal strategy under quadratic utilities and arbitrage is $\underline{\ell}_{A}$. This is due to the fact that the Sharpe Ratio is positively infinite, as the numerator is positive and the denominator is zero, assuming the risk-free asset is zero, which is reasonable for 90-minutes bets.

In summary, assuming quadratic utilities, the mean and variance of the returns $R_k$ alone provide both necessary and sufficient information to determine the optimal market portfolio that maximizes the Sharpe Ratio Criterion, which is a crucial aspect discussed in this section. It is important to highlight that this criterion exploits arbitrage opportunities, enabling strategies with no risk at all. Furthermore, the allocation between the portfolio and the risk-free asset $R_f$ can range from 0\% to 100\% of the total wealth, depending on the individual's risk tolerance. However, it is important to note that, when investing the entire budget in such a portfolio, the probability of ruin becomes positive which is a downside of this model. 

\subsection{Kelly Criterion}
The Kelly Criterion is a strategy aimed at maximizing long-term wealth by effectively balancing the potential for large returns with the risk of losses \cite{thorp1966beat}. The formula for determining the optimal strategy, denoted as $\ell_{*}$, is derived as follows.

\subsubsection{Classical Bivariate Kelly Criterion}
Consider a sequence of random rewards $\left\{R_{j}\right\}_{j}^{n}$, and let $W_{n}$ denote the final wealth of an individual who reinvests their returns according to a fixed strategy $\ell$. At time $n$, the individual's wealth is given by $W_n = W_{0} \prod_{j}^{n}R_{j}(\ell)$, where $W_{0}$ represents the initial wealth. By defining $G_{n} := W_{n}/W_{0}$ and taking logarithms, obtain the random walk expression  $G_{n} = \sum_{j}^{n}\log{(R_{j}(\ell))}$, which exhibits a drift term equal to the expected value $\mathbb{E}[\log{R_{j}(\ell)}]$. In the other side, if $S_n$ is the number of victories at time $n$ then $S_{n} \sim \text{binomial}(n, p)$. Hence, the following relationship\nospacecolon
\begin{equation*}
    \begin{split}
        W_n &=\underbrace{(1 + (\ou - 1) \ell)^{S_{n}}}_{\text{Winnings}}\underbrace{(1 - \ell)^{n - S_{n}}}_{\text{Losses}} W_0,  \\
        \iff G_{n} &= S_{n} \log{(1 + (\ou - 1) \ell)} + (n - S_{n})\log{(1 - \ell)}.
    \end{split}
\end{equation*}
Since $G_{n}$ is a sum of independent and identically distributed (i.i.d.) random variables $\log(R_{j}(\ell))$,  according to Borel's Law of Large Numbers \cite{slln_borel},
\begin{equation} \label{eq2:log_growth}
     \lim_{n \to \infty} \frac{1}{n} G_{n}(\ell) = p \log{(1 + (\ou - 1) \ell)} + (1 - p) \log{(1 - \ell)}, \quad {\text{with probability 1.}}
\end{equation}
The expression (\ref{eq2:log_growth}) denoted as to $G(\ell)$ is defined as the \textit{wealth log-growth rate} by John Kelly \cite{kelly2011new}. Since $G$ is a function of the strategy $\ell$, taking the derivative of $G$ with respect to $\ell$ eads us to the optimal solution. Thus,
\begin{equation} \label{eq2:kelly_criterion}
    G'(\ell_{*}) = 0 \iff \ell_{*} = \frac{\ou p - 1}{\ou - 1}.
\end{equation}
Recalling that in fair gambling (\ref{sct:1.1MercadoApuestas}), the odds of the event $e \in E$ are the reciprocal of the probabilities of this events. However, if $1/\ou = \tilde{p} \neq p$, hence $\ell_{*} = (p-\tilde{p})/(1-\tilde{p})$. Rearranging terms, it is obtained,
\begin{equation} \label{eq2:kelly_dkl}
    G(\ell_{*}) = p \log \left (1 + (\ou - 1)\frac{p - \Tilde{p}}{1 - \Tilde{p}} \right) + (1-p) \log \left(1 - \frac{p - \Tilde{p}}{1 - \Tilde{p}} \right) 
    = p \log \left(\frac{p}{\Tilde{p}} \right) + (1-p) \log \left(\frac{1 - p}{1 - \Tilde{p}} \right) 
    = \dkl{p}{\Tilde{p}}.
\end{equation}
Thus, the maximum log-growth is equal to the KL-Divergence (\ref{eq:1kld}). Therefore, the greater the disparity between the odds and the actual probability observed by the bettor, the greater the competitive advantage.

\subsubsection{Properties} \label{2ss:properties}
The Kelly Criterion possesses several noteworthy properties that contribute to its significance and effectiveness in optimizing long-term wealth accumulation.

Firstly, the log-growth rate $G(\ell)$ associated with the Kelly Criterion exhibits a unique optimal strategy, as established by Eduard Thorp in his paper "Optimal gambling systems for favorable games" \cite{thorp1969optimal}.There exists a critical threshold $\ell_{c} > \ell_{*}$, where $\ell_{*}$ represents the Kelly Criterion strategy, such that $G(\ell)$ transitions from positive to negative, reaching a value of zero. This property remarks emphasizes the distinct nature of different strategies in relation to their alignment with this critical threshold $\ell_{c}$.

Thorp's research \cite{thorp1969optimal} also underscores the significant impact of the chosen strategy on capital growth. When $G(\ell) > 0$, the wealth $W_{n}$ grows infinitely with probability 1, highlighting the potential for substantial wealth accumulation. Conversely, for $G(\ell) < 0$, $W_{n}$ converges to zero over time. In the case where $G(\ell) = 0$, the wealth exhibits interesting behavior, with the upper limit $\lim\sup W_{n}$ tending to infinity and the lower limit $\lim\inf W_{n}$ approaching zero as the investment horizon extends indefinitely. These findings demonstrate the sensitivity of capital growth to the chosen strategy and its profound influence on long-term financial outcomes.

Additionally, the superiority of the Kelly Criterion strategy $\ell_{*}$ over alternative strategies $\ell$ is established by Breiman \cite{breiman1961optimal}. Irrespective of the specific alternative strategy employed, portfolios adhering to the Kelly Criterion consistently outperform other strategies with probability one in terms of wealth accumulation. As the investment horizon extends indefinitely, the wealth $W_{n}(\ell_{})$ of a portfolio following the Kelly Criterion experiences infinite growth relative to the wealth $W_{n}(\ell)$ of portfolios employing alternative strategies. That is to say $\lim W_{n}(\ell_{*})/W_{n}(\ell) = \infty$ when $n \to \infty$, with probability 1. This highlights the remarkable advantage of the Kelly Criterion in maximizing long-term wealth accumulation.

The properties of the Kelly Criterion underscore its significance and effectiveness in optimizing long-term wealth accumulation. The distinct nature of strategies in relation to the critical threshold, the sensitivity of capital growth to the chosen strategy, and the superior performance of the Kelly Criterion strategy over alternatives all contribute to its importance. By aligning with the Kelly Criterion, investors can enhance their wealth accumulation potential, leading to more favorable financial outcomes in the long run.

\subsubsection{Multivariate Kelly Criterion}
Expanding Kelly's Criterion to encompass multivariate bets, where there are $m$ possible events associated with the same phenomenon, when event $e_{i}$ occurs the raw return is denoted as $R(e_{i}; \underline{\ell}) = 1 + \ou_{i}\ell_{i} - \sum_{j}^{m}\ell_{j}$, where $\underline{\ell}$ represents the vector of strategies corresponding to each event.

Considering $S_i$as the number of occurrences of the $i$-th outcome where $\sum_{j}^{m}S_j = n$, then the wealth at trail $n$ is given by $W_n(\underline{\ell}) = \prod_{i}^{m}\left (1 + \ou_{i}\ell_{i} - \sum_{j}^{m}\ell_{j} \right)^{S_{i}}W_0$. By taking logarithms and considering the limit as $n$ approaches infinity, the log-growth rate is obtained as followed \cite{smoczynski2010explicit}\nospacecolon
\begin{equation*}
    G(\underline{\ell}) = \sum_{i = 1}^{m} p_{i} \log \left(1 + \ou_{i}\ell_{i} - \sum_{j=1}^{m}\ell_{j} \right), \quad \text{with probability 1.}
\end{equation*}

The formulation of the Multivariate Kelly Criterion in a matrix-based representation constitutes an original contribution. By introducing the probability vector $\underline{p} \in [0,1]^{m}$ as the vector of probabilities associated with each event and the consequences matrix $W = \left[\underline{w}_{1}|\underline{w}_{2}|\dotsc|\underline{w}_{m}\right]$, where $\underline{w}_{j} = \ou_{j}\hat{\mathbf{e}}_{j}$ and $\hat{\mathbf{e}}_{j}$ is the $j$-th canonical vector, the problem of determining the Multivariate Kelly Criterion can be cast. The objective is to maximize the expression 

\begin{equation} \label{eq2:multiple_kelly}
	\maxi \left \{\underline{p}' \log\left(\mathbf{1} + W'\underline{\ell} - \sum_{i=1}^{m}\ell_{i} \mathbf{1} \right)\right \} \quad \text{subject to} \quad \sum_{i=1}^{m}\ell_{i} \leq 1, \quad \underline{\ell} \geq \mathbf{0}. 
\end{equation}

Importantly, the optimization problem associated with the Multivariate Kelly Criterion exhibits a concave structure, enabling its solution through convex optimization algorithms like Successive Quadratic Programming (SQP) \cite{busseti2016risk}. While Smoczynski has proposed an algorithm for determining $\underline{\ell}_{*}$ in his work \cite{smoczynski2010explicit}, a closed-form solution is not available. The formulation of this criterion in matrix form significantly contributes to a deeper comprehension of its characteristics. By focusing on the functional form of the log growth and obtaining analytical gradients, we can effectively maximize the criterion function. Furthermore, this matrix-based approach facilitates practical implementation by eliminating the need for iterative loops and relying solely on linear algebraic operations. As a result, this formulation not only enhances theoretical understanding but also provides valuable insights for the efficient application of the Multivariate Kelly Criterion.

\subsubsection{Multivariate and Simultaneous Kelly Criterion}
In the most general case, the multivariate criterion has been extended to incorporate simultaneous random rewards. This extension represents a novel development in the field. In this scenario, a set of $r$ independent random rewards occurs simultaneously, each with $m_{k}$ possible events. The probability space is defined as $\Omega = \bigtimes_{k}^{r}\Omega_{k}$, resulting in a total of $M = \sum_{k}^{r} m_{k}$ events and $N = \prod_{k}^{r} m_{k}$ possibilities. The strategy vector is defined as the concatenation of the betting vectors of the $r$ random rewards, denoted as $\underline{\ell} = (\underline{\ell}_{1}, \underline{\ell}_{2}, \dots, \underline{\ell}_{r})' \in \R^{M}$. Similarly, the decimal odds vector is represented as $\underline{\ou}$ and the net returns vector as $\underline{\varrho}$. The overall raw return of the strategy, denoted as $R(\underline{\ell})$, can be expressed as $R(\underline{\ell}) = 1 + \underline{\ell}'\underline{\varrho}$. The matrix of consequences, denoted as $W \in \R^{M \times N}$, represents the profit possibilities, with each column corresponding to a specific profit outcome $\underline{\omega}_{j_{k}} = \underline{\omega}_{j}$ for each one of the sample spaces $\Omega_{k} \subseteq \Omega$.
\begin{equation*}
    \text{i.e.} \quad \underline{w}_{j} = D_{\ou} (\hat{\mathbf{e}}_{j_{1}}, \hat{\mathbf{e}}_{j_{2}}, \dots, \hat{\mathbf{e}}_{j_{r}})', \quad j_{k} \in \underline{\omega}_{j}, \forall k = 1, 2, \dots, M.
\end{equation*}
Since the outcomes of each bet are assumed to be independent, the probability vector is given by 
\begin{equation} \label{eq2:probas_simult}
    \underline{p}_{i} = \mathbb{P}\left[\underline{\varrho}(\underline{\omega}_{i}) \right] = \mathbb{P}\left[ \bigcap_{k}^{r}\underline{\varrho}_{k}(\omega_{i,k}) \right] = \prod_{k}^{r} \mathbb{P}\left [\underline{\varrho}_{k}(\omega_{i,k}) \right].
\end{equation}
Consequently, the Multivariate and Simultaneous Kelly Criterion is formulated as the optimization problem\nospacecolon
\begin{equation} \label{eq2:multiple_simult_kelly}
	\maxi \left \{\underline{p}' \log\left(\mathbf{1} + W'\underline{\ell} - \sum_{i=1}^{M}\ell_{i} \mathbf{1} \right)\right \} \quad \text{subject to} \quad \sum_{i=1}^{M}\ell_{i} \leq 1, \quad \underline{\ell} \geq \mathbf{0}. 
\end{equation}

The extension of the Multivariate Kelly Criterion to incorporate simultaneous and multiple random rewards presents significant advantages and opportunities for optimizing decision-making in intricate betting scenarios. As previously discussed in the context of the Multiple Kelly Criterion, this expanded framework offers a distinct advantage by enabling analytical and programmatically efficient implementation. By leveraging the functional form of the criterion, decision-makers can employ a numerical approach that is both robust and stable within the realm of matrix algebra. 

The aforementioned properties of the Kelly Criterion highlight its resilience and competitive edge in the pursuit of long-term wealth accumulation. The existence of a distinctive optimal strategy, coupled with the profound influence of strategy selection on capital growth, solidifies the Criterion's efficacy and prominence within the realm of financial decision-making. Building upon these foundational principles, we can now delve into a real-world application that pits the Kelly Criterion against the Sharpe Criterion: betting in the English Premier League. This practical demonstration will further illuminate the practical implications and comparative performance of these two criteria in a tangible and relevant context.

\section{Results}

This section presents the optimal betting portfolios for the second part of the 2020/2021 season of the English Premier League. These portfolios are derived using the Sharpe Ratio Criterion and the Kelly Criterion based on odds estimated by deep learning models. For each criterion, two types of strategies were examined\nospacecolon \ unrestricted strategies and restricted strategies. The former allows for betting on all possible events of each match, while the latter imposes limitations on the number of events that can be bet upon.

It is worth noting that excessive betting can lead to a negative log-growth rate $G(\underline{\ell}) < 0$, resulting in the wealth tending towards zero, as mentioned in the first property of the Kelly Criterion \cite{thorp1969optimal}.  However, according to the findings of E. Thorp stated at \ref{2ss:properties}, this negative impact occurs after reaching the optimal point. Therefore, it is preferable to underestimate the bets rather than overestimate them. This rationale underscores the inclusion of fractional bets in the strategies.

\subsection{Predictive Model}
The predictive model was constructed based on data obtained from three distinct public sources. Firstly,  \href{https://sofifa.com/}{EA Sports' ratings} \cite{sofifa} were utilized to assess the overall quality, offense, midfield, and defense of teams on a weekly basis. Secondly, team statistics were collected from \href{https://understat.com/}{Understats} \cite{understat}. Lastly, match odds from multiple bookmakers, obtained through \href{https://www.football-data.co.uk/englandm.php}{Football Data U.K.}. \cite{footballdata}, were integrated into the model.

\subsubsection{Data}
The dataset encompassed the period from the 2014/2015 season to the 2020/2021 season. To ensure the reliability of the analysis, data from the initial week of each season was excluded. These initial weeks were deemed less informative due to ongoing team rebuilding during the summer transfer market, introducing substantial uncertainty in team performance.

The final dataset comprised a total of 2,660 games, which were divided into three distinct groups\nospacecolon \ training (2014/2015 to 2019/2020 seasons), validation (first half of the 2020/2021 season), and test (second half of the 2020/2021 season). Temporal variables were aggregated using weighted maximum likelihood, employing an exponential decay rate of 0.1. This particular value was suggested by the Dixon-Coles analysis \cite{dixon1997modelling} and effectively captured the diminishing impact of historical data as time progressed. Moreover, the model considered variables for both home and visiting teams, encompassing various aspects of team performance (refer to the appendix for detailed variable information).

As an additional input to the model, the final odds from Pinnacle Sports \cite{pinnacle}, a reputable bookmaker, were incorporated. This incorporation was motivated by the by the work of Surowiecki, which recognize that market information holds predictive power in estimating potential events \cite{surowiecki2005wisdom}. By integrating these odds, the model could effectively leverage market insights to enhance its predictive capabilities.

\subsubsection{Model Selection}
To determine the variables with the highest predictive power, lasso-shrinkage techniques were employed for a multinomial regression analysis \cite{tibshirani1996regression}. The use of variable selection was motivated by the relatively limited size of the dataset compared to the number of parameters in the neural network, which increased the risk of overfitting.

By employing the lasso technique, a subset of variables with the most significant predictive impact was identified. The multinomial regression model trained on this reduced set of variables achieved a predictive accuracy of 41.76\% and a cross-entropy value of 1.34. In contrast, when trained on the complete set of variables, the model achieved an accuracy of 51.76\% and a cross-entropy value of 1.22. It is important to note that the coefficients of the multinomial regression model were not easily interpretable, and the accuracy was significantly reduced by 19.3\% compared to the original model, while the entropy was 9.8\% higher. Considering these findings, it was determined that utilizing all variables in the subsequent deep learning models would be more beneficial. This decision was influenced by the understanding that deep learning models have the capability to effectively capture complex relationships and patterns present in the data, even if certain variables may have less intuitive interpretations.
\begin{figure}[ht]
\begin{subfigure}{.5\textwidth}
  \centering
  \includegraphics[width=1\linewidth]{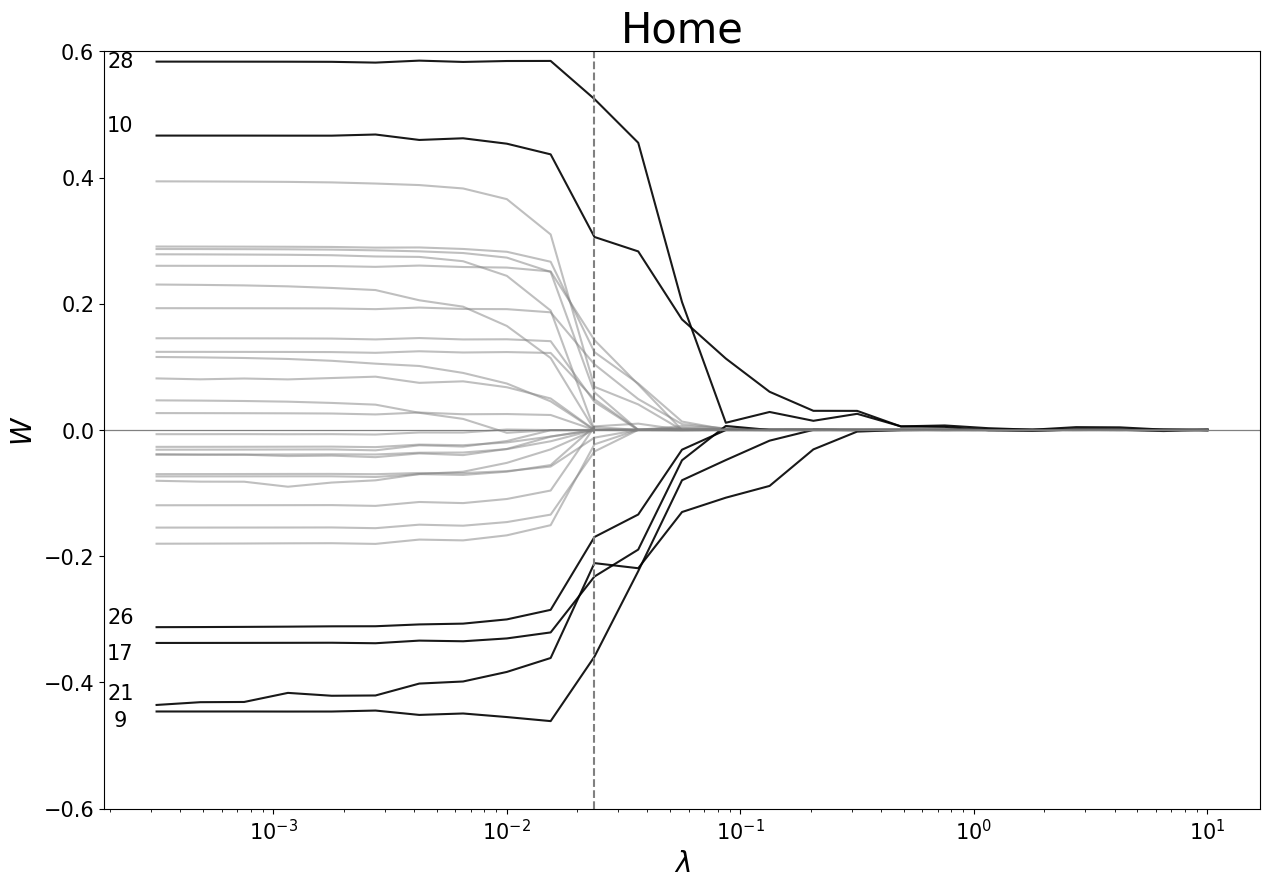}
  \label{fig:3.0.1}
\end{subfigure}%
\begin{subfigure}{.5\textwidth}
  \centering
  \includegraphics[width=1\linewidth]{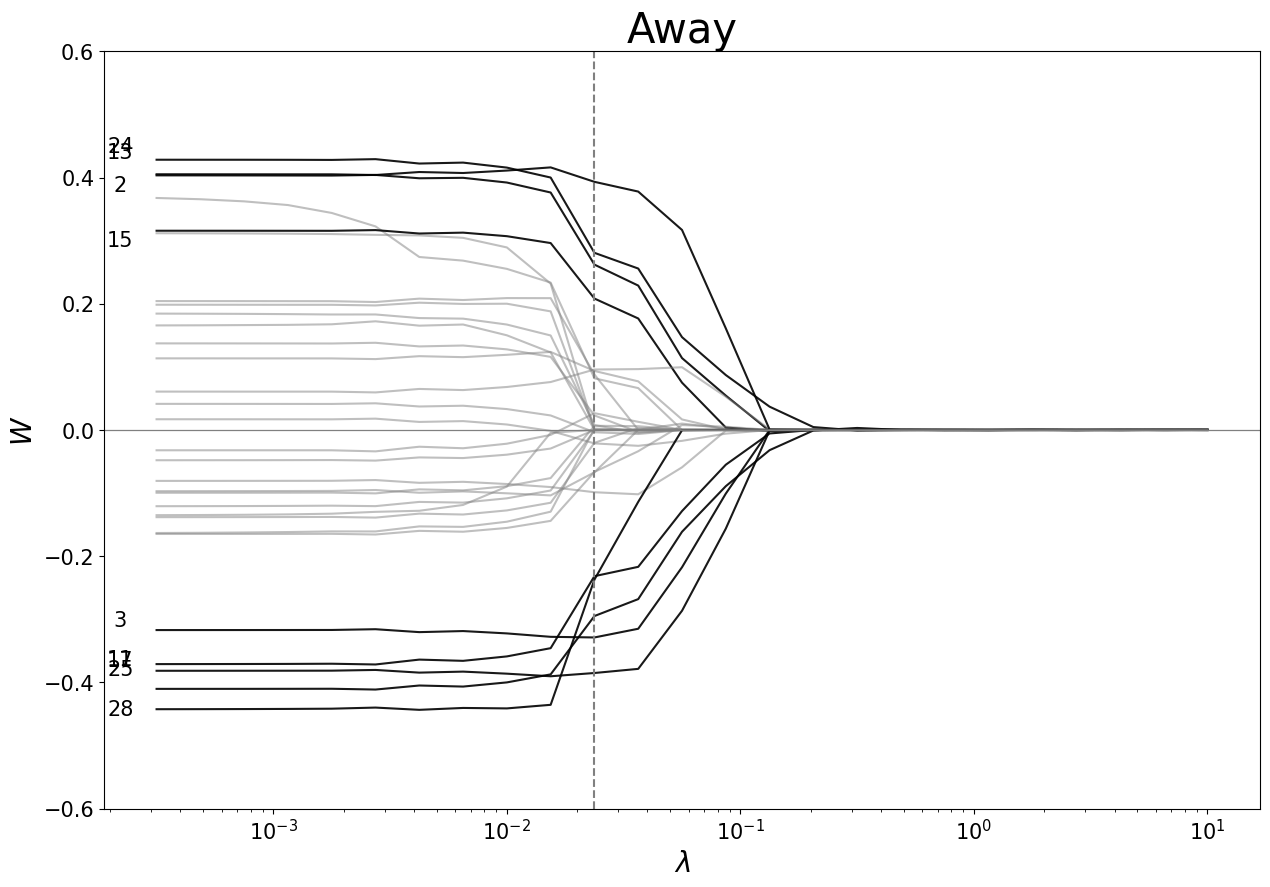}
  \label{fig:3.0.3}
\end{subfigure} %
\caption{Graphs illustrating the decay of weights in the lasso penalization models for the \texttt{Home} and \texttt{Away} outcome. The bold lines correspond to variables whose absolute coefficients exceed one standard deviation of all coefficient values. Detailed information on the specific variables and their associated numbers can be found in the appendix.}
\label{fig:lasso_pen}
\end{figure}

A training set consisting of 2,200 observations was obtained, while the number of parameters to estimate was 2,700, given the $30 \times 30 \times 3$ dimensions. Consequently, regularization was necessary to address the challenge of having more variables than observations.

Furthermore, certain hyperparameters were fixed across all deep learning models with different regularizations using the framework described at \cite{nielsen2015neural} \cite{geron2019hands}. These fixed hyperparameters included the architecture design, where a funnel architecture was employed, progressing from a higher number of neurons to a lower number as the number of hidden layers increased. The random seed and kernel initializer were also standardized, using He-Normal for hidden layers and Glorot-Normal for the output layer \cite{geron2019hands}. The number of hidden layers was set to three, and for models incorporating batch normalization regularization, the number of hidden layers became a hyperparameter. Additionally, the NAdAM learning algorithm \cite{nocedal1999numerical} was utilized consistently across all models.

Subsequently, several hyperparameters were learned through the optimization process. The number of neurons in the first layer was explored within a range of 70\% to 200\% of the number of variables, while for subsequent layers, the algorithm initiated the search with the number of neurons from the previous layer. The learning rate was selected from a list of three values\nospacecolon \ $10^{-2}$, $10^{-3}$, and $10^{-4}$. The penalization magnitude and the convexity trade-off (elastic net) were determined by sampling penalty values from a uniform distribution.
\begin{table}[ht]
\captionsetup{font=footnotesize}
\centering
\resizebox{16cm}{!}{
    \begin{tabular}{@{}lccccc@{}}
    \toprule
    \multicolumn{1}{c}{\textbf{Hyperparameters \textbackslash \ Model}} & \textbf{Elastic Net} & \textbf{Lasso} & \textbf{Ridge} & \textbf{Drop Out} & \textbf{Batch Norm} \\ \midrule
    \textbf{Num. Layers}                                        & 3                    & 3              & 3              & 3                 & 9                   \\
    \textbf{Num. Neurons}                                     & 37, 25, 21           & 25, 21, 25     & 37, 33, 33     & 25, 41, 37        & 33,33,37,37,37,37,49,45,97                \\
    \textbf{Learning Rate}                                  & 0.01                 & 0.01           & 0.01           & 0.001             & 0.01                \\
    \textbf{Penalty}                                     & 0.9, 0.8, 0.1, 0.3           & 0.1, 0.5, 0.06, 0.5   & 0.6, 0.8, 0.2, 0.01   & -                 & -                   \\
    \textbf{Convexity}                                      & 0.12, 0.03, 0.33, 0.23        & -              & -              & -                 & -                   \\
    \textbf{Dropout Rate}                                   & -                    & -              & -              & 0.1122, 0.00025, 0.000055   & -                   \\ \bottomrule
    \end{tabular}
}
\caption{\label{tab:table_model_stats} Optimal Hyperparameters for Different Regularizations in the Validation Set.
}
\end{table}

To evaluate the predictive errors and information loss, cross-temporal validation techniques, specifically the Split Temporal Cross-Validation (STCV) approach, were employed.

\begin{table}[ht]
\captionsetup{font=footnotesize}
\centering
\resizebox{8cm}{!}{
    \begin{tabular}{@{}ccccc@{}}
    \toprule
    \multicolumn{1}{c} \textbf{\textbf{Elastic Net}} & \textbf{Lasso} & \textbf{Ridge} & \textbf{Drop Out} & \textbf{Batch Norm} \\ \midrule
    1.02                                       & 1.03                   & 1.01            & 1.99             & 1.02                               \\
    \bottomrule
    \end{tabular}
}
\caption{\label{tab:table_model_entropy} Cross Entropy in the Validation Set.
}
\end{table}

\subsubsection{Model Assessment}
Subsequently, the best models selected from the training and validation sets were trained. Among the different architectures considered, the \textit{Drop Out} architecture \textit{consistently} demonstrated superior performance in terms of precision and information loss. Therefore, it was chosen as the best model for further analysis.

\begin{table}[ht]
\captionsetup{font=footnotesize}
\centering
\resizebox{6cm}{!}{
\begin{tabular}{@{}lcc@{}}
    \toprule
    \multicolumn{1}{c}{\textbf{Model}} & \textbf{Loss}    & \textbf{Accuracy} \\ \midrule
    \textbf{Elastic Net}                & 1.0288 $\pm$ 0.1241 & 58\% $\pm$ 16.91\%    \\
    \textbf{Lasso}                      & 1.0405 $\pm$ 0.1234    & 55\% $\pm$ 16.88\%    \\
    \textbf{Ridge}                      & 1.0185 $\pm$0.1273     & 57\% $\pm$ 19\%       \\
    \textbf{Drop Out}                   & 0.9819 $\pm$0.0996     & 55\% $\pm$11.61\%     \\
    \textbf{Batch Norm}                 & 0.9386 $\pm$0.1569     & 57\% $\pm$ 17.64\%    \\ \bottomrule
    \end{tabular}
}
\caption{\label{tab:table_models_metrics} The average of the metrics plus/minus one sample standard deviation, in the test set is presented using STCV.}
\end{table}

The predictions made by the model were compared to the pre-match odds provided by the bookmaker Pinnacle Sports. This comparison was motivated by the belief that the odds reflect the collective wisdom of the crowds \cite{surowiecki2005wisdom}.
In terms of accuracy, the model achieved a 54\% accuracy rate, while the crowds achieved 51.5\%. In comparison, the null models of "always bet on the home team," "always bet on a draw," and "always bet on the away team" had accuracy rates of 38\%, 20\%, and 42\% respectively.
Regarding information loss, the model achieved a value of 1.0318, while the crowds achieved 0.9966, and the historical frequencies up to the 19th matchweek of the 2020/2021 season had an information loss value of 1.0631.

\begin{table}[ht]
\captionsetup{font=footnotesize}
\centering
\resizebox{9cm}{!}{
\begin{tabular}{ll|ccc|c}
\multicolumn{2}{c|}{\multirow{2}{*}{\diagbox{\textbf{Drop Out}}{\textbf{Market Odds}}}} & \multicolumn{4}{c}{\textbf{Prediction}}                                    \\
\multicolumn{2}{c|}{}                                                                     & Home                 & Draw     & Away               & \textbf{Total}  \\ 
\hline
\multirow{4}{*}{\textbf{Observed}} & Home                                                & 60\textbackslash{}55 & 0\textbackslash{}0          & 16\textbackslash{}21 & \textbf{76}     \\
                                    & Draw                                              & 17\textbackslash{}15 & 0\textbackslash{}0          & 23\textbackslash{}25 & \textbf{40}     \\
                                    & Away                                              & 36\textbackslash{}36 & 0\textbackslash{}0         & 48\textbackslash{}48 & \textbf{84}     \\ 
\cline{2-6}
                                    & \textbf{Total}                                      & \textbf{113\textbackslash{}106}         & \textbf{0\textbackslash{}0} & \textbf{87\textbackslash{}94}          & \textbf{200}   
\end{tabular}
}
\caption{\label{tab:mat_conf} The confusion matrix compares the predictions of the models to the actual outcomes. The lower part of the table corresponds to the predictions made by the models, while the last column represents the observed outcomes. The elements on the left side of the diagonals correspond to the predictions made by the Drop Out model, while the elements on the right side represent the odds ratios provided by Pinnacle Sports.}
\end{table}

\subsection{Portfolios} 

This section presents the results of the optimal portfolios based on quadratic and logarithmic utilities. It is important to note the following assumptions and constraints throughout the analysis\nospacecolon \ no more than 100 percent of the wealth can be bet, short selling is not allowed, the risk-free asset has a zero return, money is infinitely divisible, the portfolio is also infinitely divisible\footnote{For numerical purposes, any wager less than 0.0001 was considered negligible.}. Without loss of generality the initial wealth is set to \$1. Gains are reinvested each matchweek, i.e. $W_{n} = \prod_{\text{J}19}^{\text{J}38}R_{i}(\underline{\ell})$.

Four types of scenarios were considered for the portfolios. Each strategy was evaluated under two scenarios\nospacecolon \ restricted betting, where only a single event per match can be bet upon, and unrestricted betting, allowing for multiple bets per match. Furthermore, portfolios were examined for both full strategies and fractional strategies at $f = $17\%. The optimal fraction of 17\% for this set was determined after conducting two hundred Dirichlet simulations in the validation set. In practice, it is common to use 25\% of the Kelly Criterion strategy for betting purposes. The decision to split the bets is based on the insight that under-betting is preferable to over-betting, as indicated by E. Thorp \cite{thorp1969optimal}.
\begin{figure}[ht]
    \begin{center}
        \includegraphics[scale=0.20]{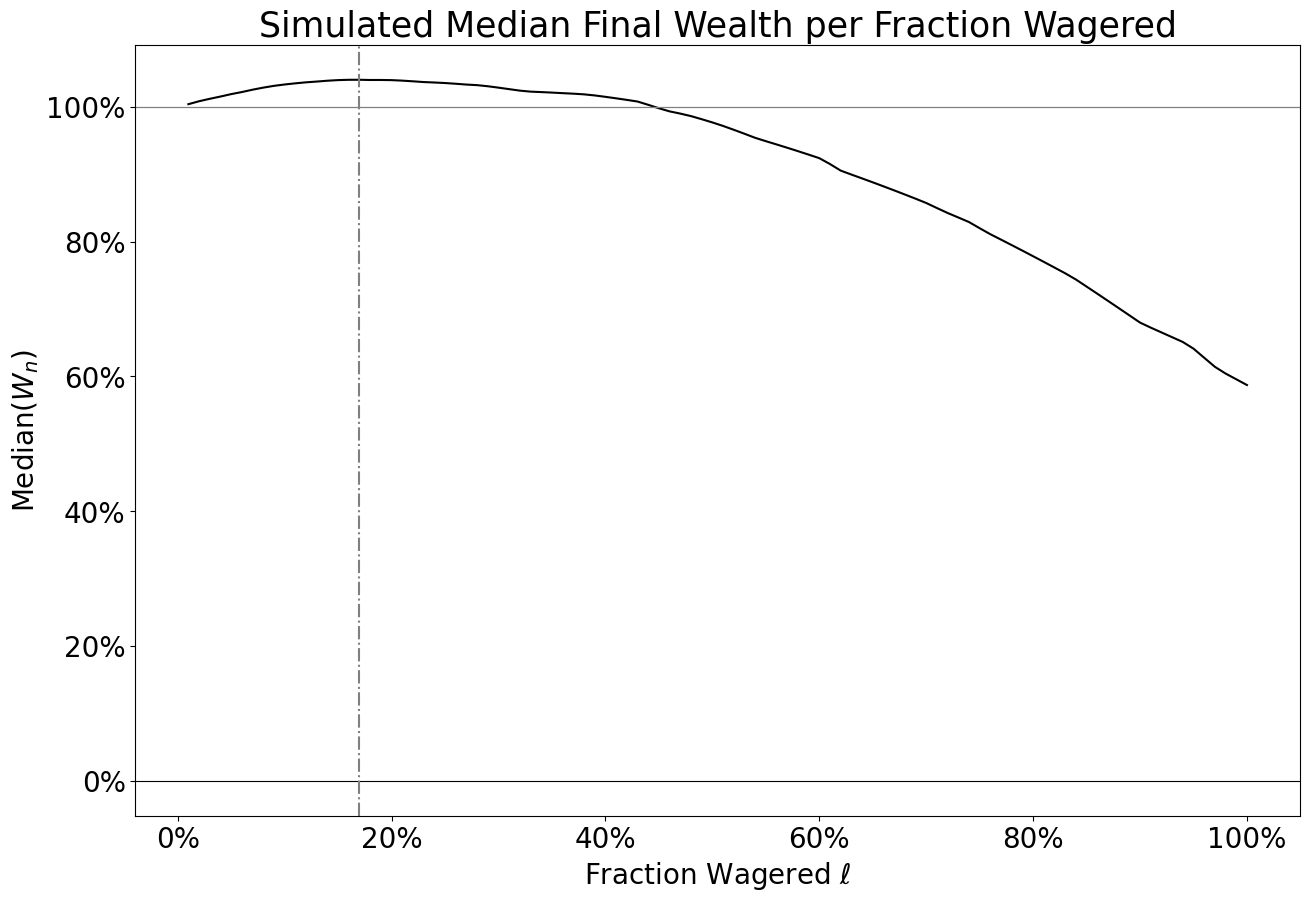}
        \caption{Plot of the median final wealth curve obtained by two hundred $\text{Dirichlet}(\mathbf{1})$ simulations on the validation set.}
        \label{fig:acc_loss_all}
        \centering
    \end{center}
\end{figure}

\subsubsection{Complete Strategies}
In the test set, there are 20 matchweeks, with 10 matches taking place on each match day. Each match has three possible outcomes\nospacecolon \ home win, draw, or away win. The results of one match are independent of the results of the other matches. Consequently, there are $M = \sum_{i}^{r}m_{i} = 3 \times 10 = 30$ possible bets and $N = \prod_{i}^{r}m_{i} = 3^{10} = 59,049$ combinations of outcomes per matchweek.

For portfolios with logarithmic utilities, the optimal bets were determined by optimizing the Multiple Simultaneous Kelly Criterion (\ref{eq2:multiple_simult_kelly}) using the SQP algorithm. On average, this model took approximately 20.9 seconds to converge per fixture. However, it should be noted that the numerical algorithm did not converge for matchweek 23 due to gradient overflow. Despite this challenge, it was decided to use the stakes from the last iteration of the algorithm as an approximation for that matchweek. Although it is important to acknowledge this approximation, it was necessary in order to maintain the continuity of the analysis and ensure the inclusion of matchweek 23 in the overall assessment of the strategies.

In the case of quadratic utilities, the optimal portfolios were obtained by maximizing the Sharpe Ratio through the solution of the convex optimization problem (\ref{eq:2.min_sharperatio_convex}) using the SQP algorithm. On average, the algorithm took 5.3 seconds per fixture to converge. It is worth mentioning that all optimizations successfully converged for the Sharpe Ratio criterion.

\begin{figure}[ht]
\begin{subfigure}{.52\textwidth}
  \centering
  \includegraphics[width=0.9\linewidth]{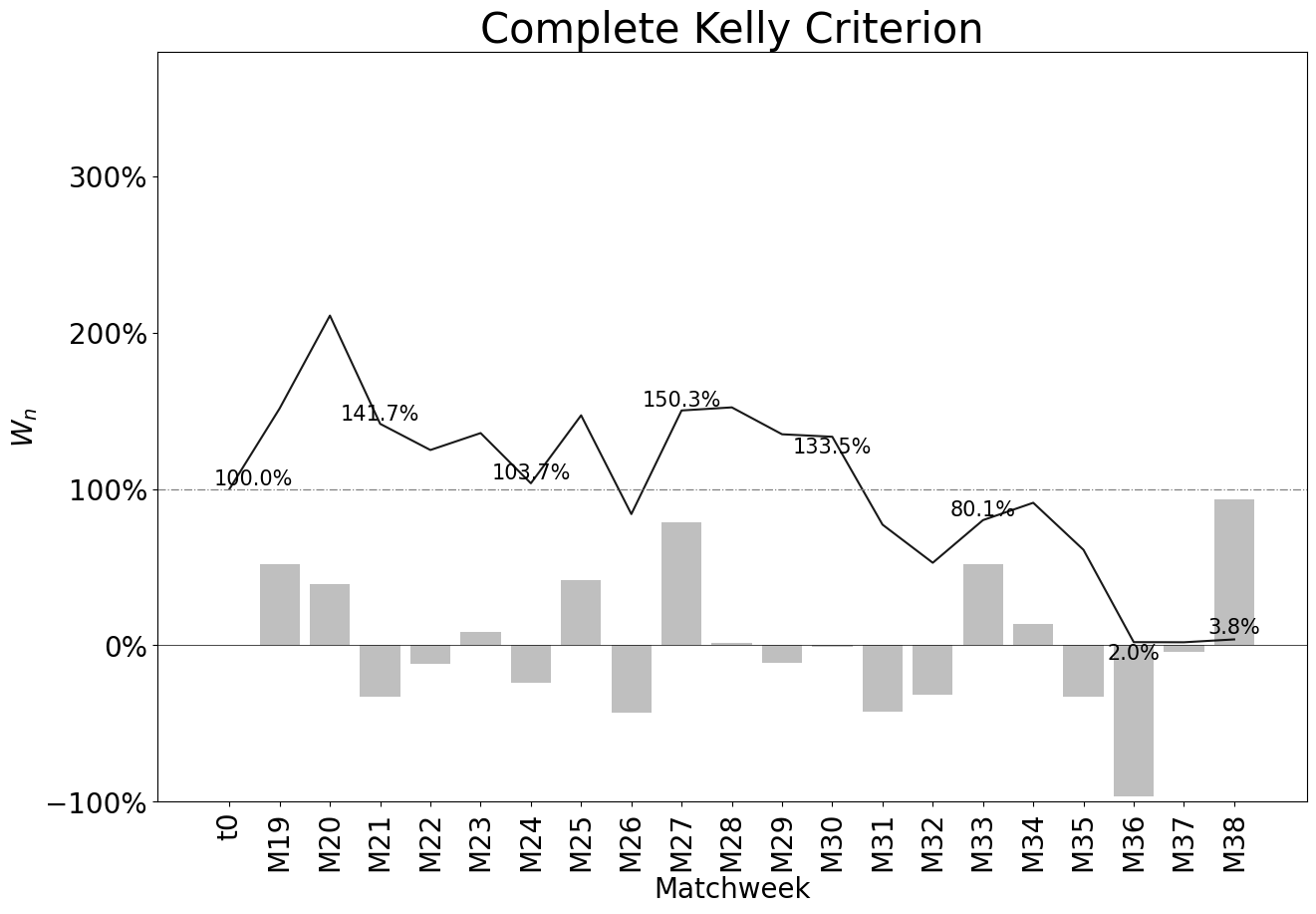}
  \label{fig:3.1.1}
\end{subfigure}%
\begin{subfigure}{.52\textwidth}
  \centering
  \includegraphics[width=0.9\linewidth]{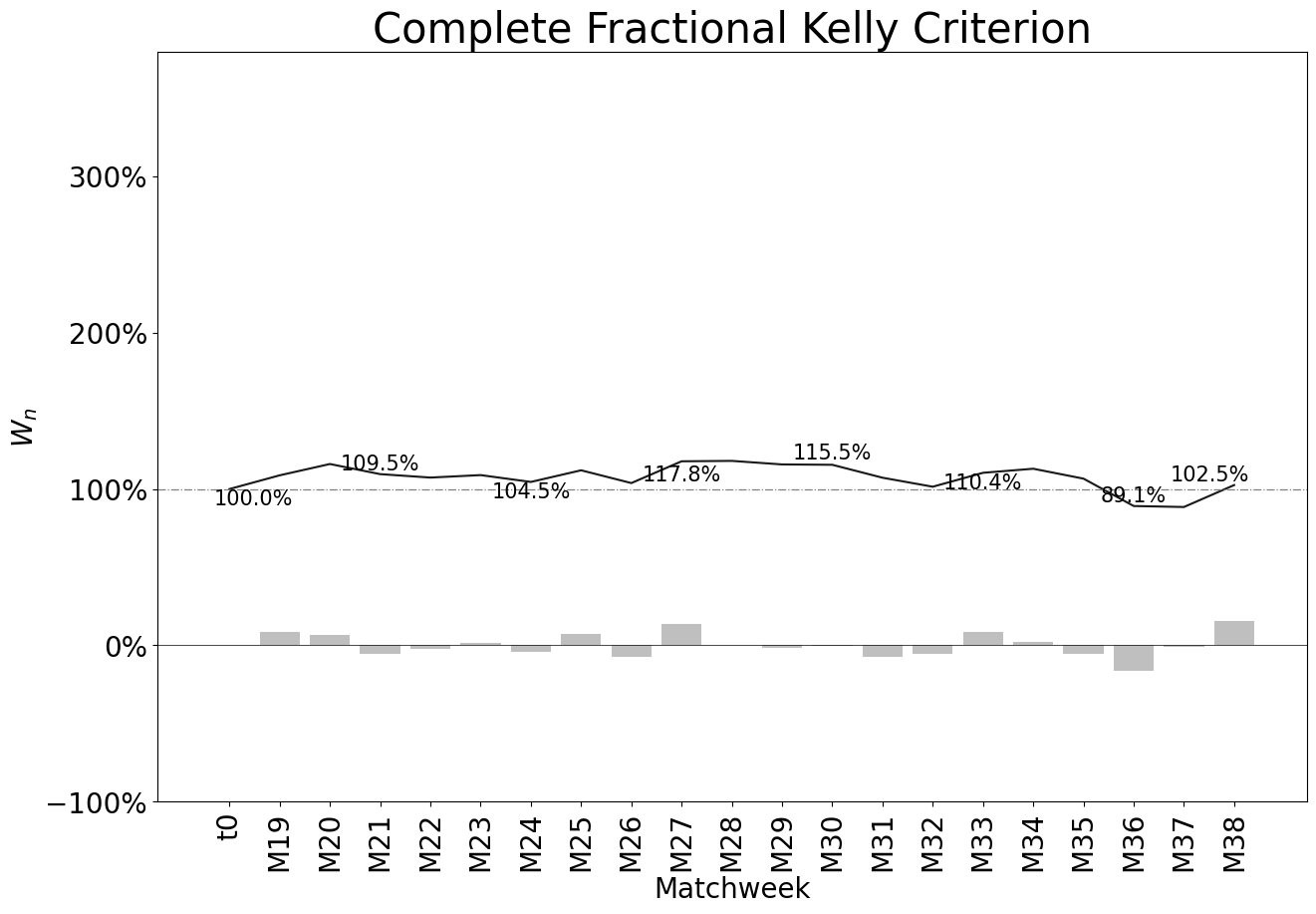}
  \label{fig:3.1.2}
\end{subfigure} %
\begin{subfigure}{.52\textwidth}
  \centering
  \includegraphics[width=0.9\linewidth]{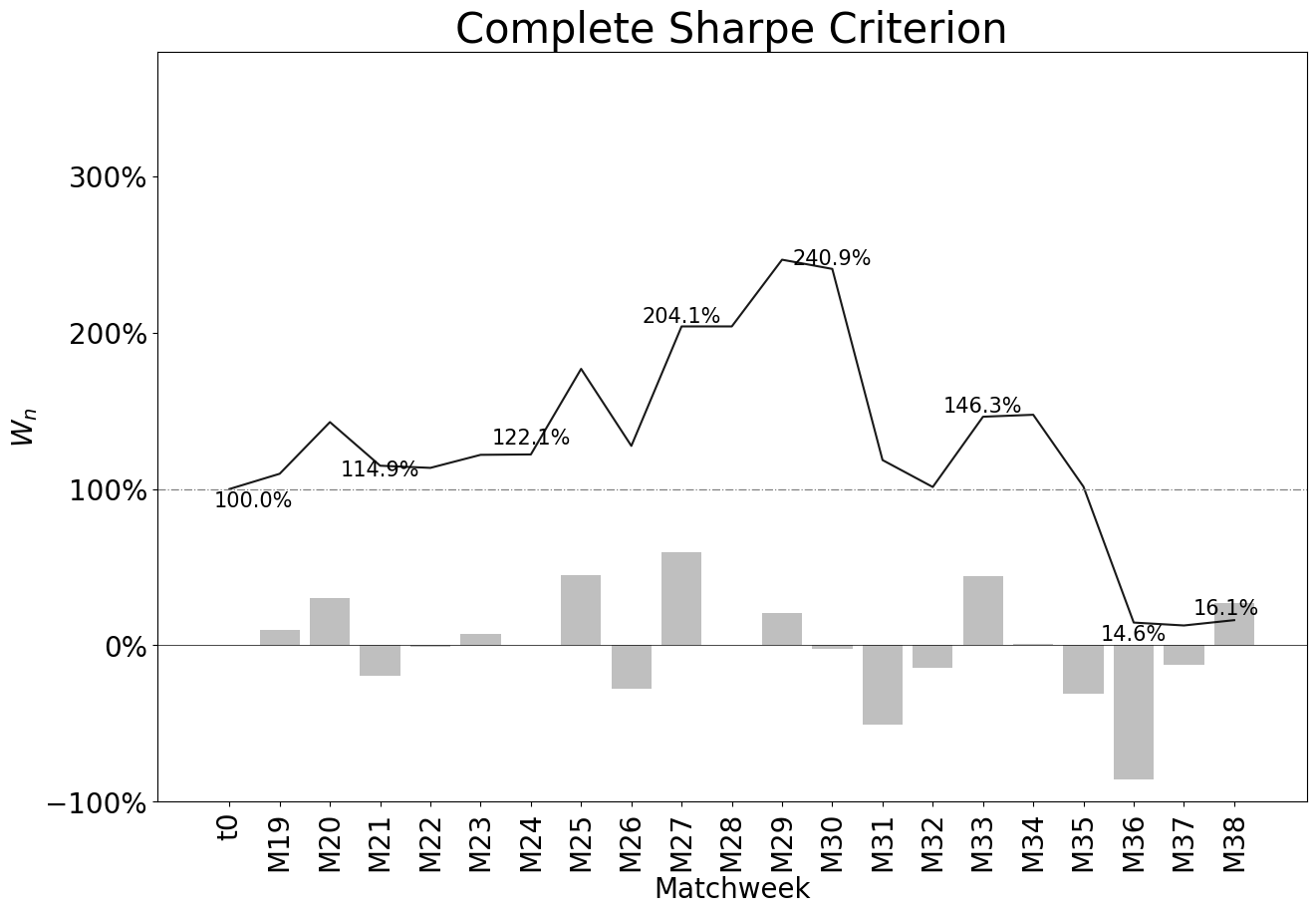}
  \label{fig:3.1.3}
\end{subfigure} %
\begin{subfigure}{.52\textwidth}
  \centering
  \includegraphics[width=0.9\linewidth]{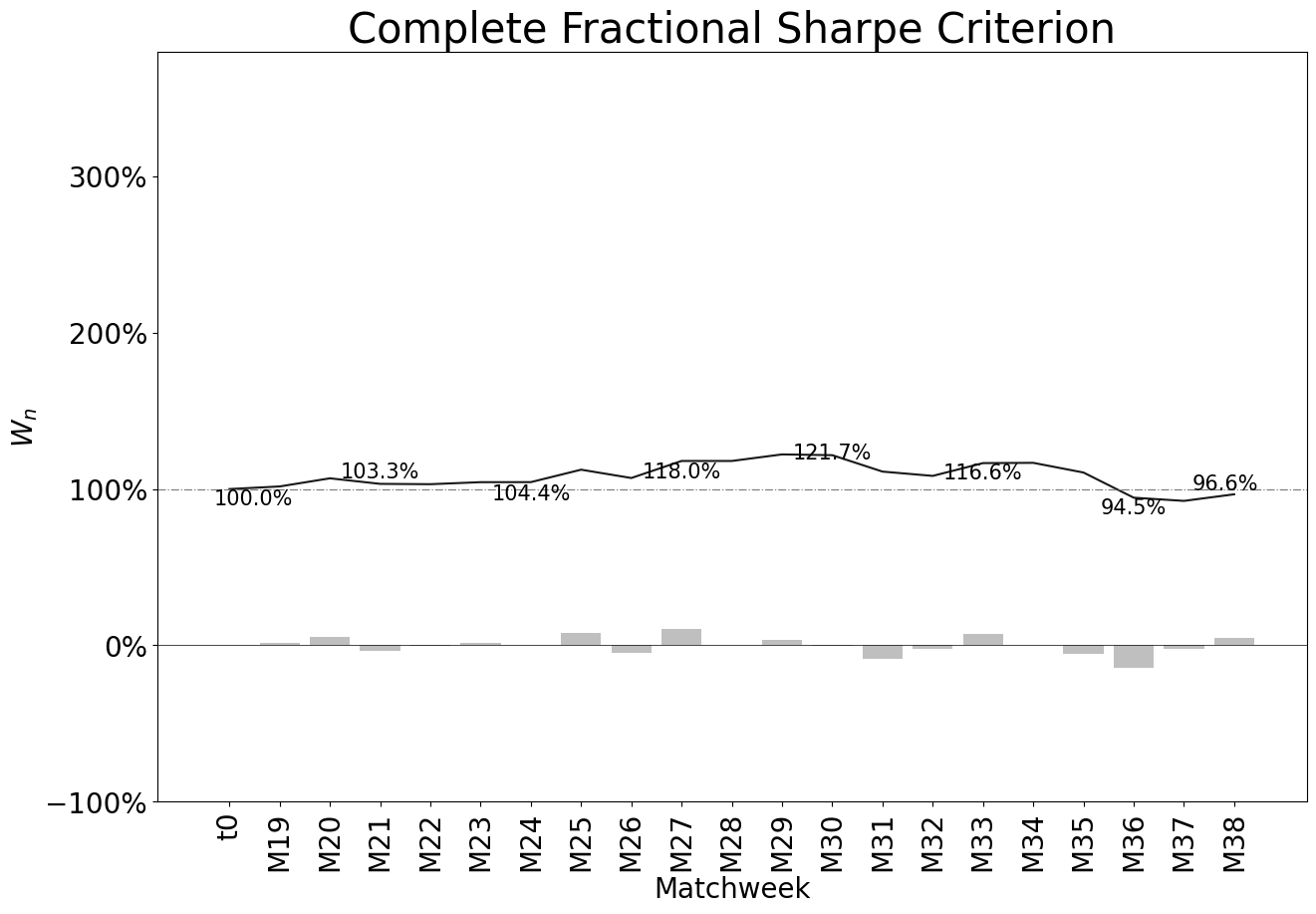}
  \label{fig:3.1.4}
\end{subfigure} %
\caption{Plots of wealth $W_{n}$ under complete strategies for both utilities. The line represents the accumulated wealth up to the $i$-the matchweek and the bars represent the profit or loss for the given matchweek.}
\label{fig:complete_models}
\end{figure}

Three matches with arbitrage opportunities across three different matchweeks were identified. One notable example is the match between Everton and Aston Villa on June 5, 2021, which had a commission of $tt = -0.0085$. The following table provides a summary of the performance of both strategies in comparison to the match featuring arbitrage on that particular match day.

\begin{table}[ht]
\captionsetup{font=footnotesize}
\centering
\resizebox{10cm}{!}{
\begin{tabular}{@{}lcccccc@{}}
    \toprule
    \multicolumn{1}{c}{\textbf{Result}} & \textbf{$\hat{p}$}  & \textbf{$\ou$} & \textbf{$\ou^{-1}$} & \textbf{$\ou^{-1} / (1 + tt)$} & \textbf{$\ell_{*}^{(S)}$} &  \textbf{$\ell_{*}^{(K)}$}  \\ \midrule
    \textbf{Home}                          & 42.34\%          & 2.07          & 48.31\%               & 48.73\%                         & 48.71\%          &0.52\%\\
    \textbf{Draw}                        & 32.08\%          & 3.7           & 27.03\%               & 27.26\%                         & 27.25\%          &7.35\%\\
    \textbf{Away}                        & 25.58\%          & 4.2           & 23.81\%               & 24.01\%                         & 24.01\%          &4.04\%\\
    \textbf{Total}                         & \textbf{100.0\%} & \textbf{-}    & \textbf{99.1\%}       & \textbf{100.0\%}                & \textbf{99.7\%} & \textbf{11.91\%}\\ \bottomrule
    \end{tabular}
}
\caption{\label{tab:table_arbitrage} Sharpe Criterion $\ell_{*}^{(S)}$ and Kelly $\ell_{*}^{(K)}$ strategies for the 2021 Everton v.s. Aston Villa match.}
\end{table}

\subsubsection{Restricted Strategies}

The restricted betting sample space consists of the event with the highest expected value. Mathematically, it can be defined as $\Omega := \bigtimes_{i}^{r} \left\{\omega_{k} \suchthat k = \argmax_{j} \{\mathbb{E}[\varrho_{i,j}]\}, \quad j = 1, 2, \dots, m_{i} \right\}$. Consequently, the number of possible outcomes is reduced to $N = 2^{10}$, and the total number of bets is $M = 3 \times 10$. The algorithms for both utilities exhibited convergence, with an average convergence time of approximately 1 second for all match days, and there were no instances of non-convergence.

\begin{figure}[ht]
\begin{subfigure}{.52\textwidth}
  \centering
  \includegraphics[width=0.9\linewidth]{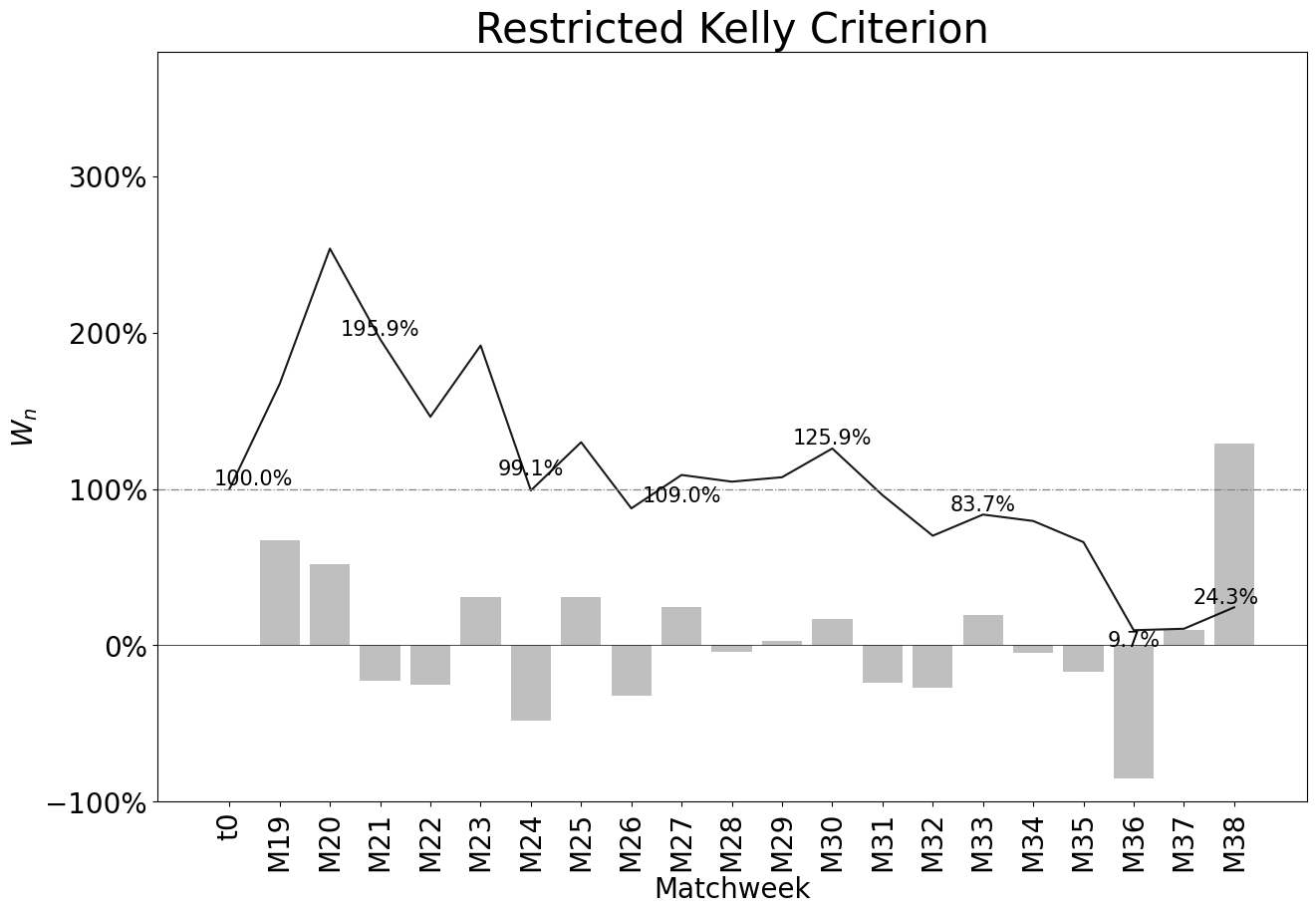}
  \label{fig:3.2.1}
\end{subfigure}%
\begin{subfigure}{.52\textwidth}
  \centering
  \includegraphics[width=0.9\linewidth]{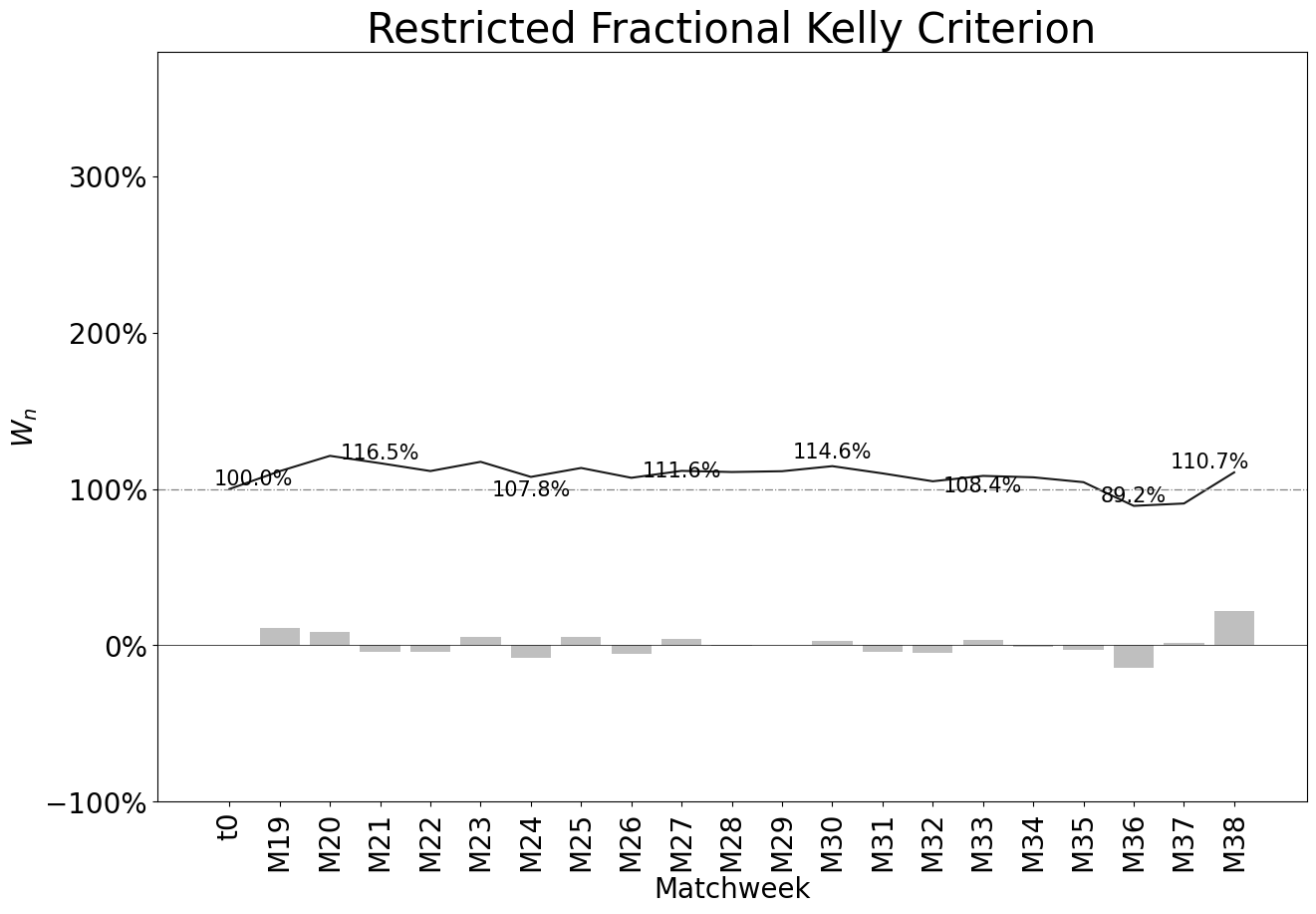}
  \label{fig:3.2.2}
\end{subfigure} %
\begin{subfigure}{.52\textwidth}
  \centering
  \includegraphics[width=0.9\linewidth]{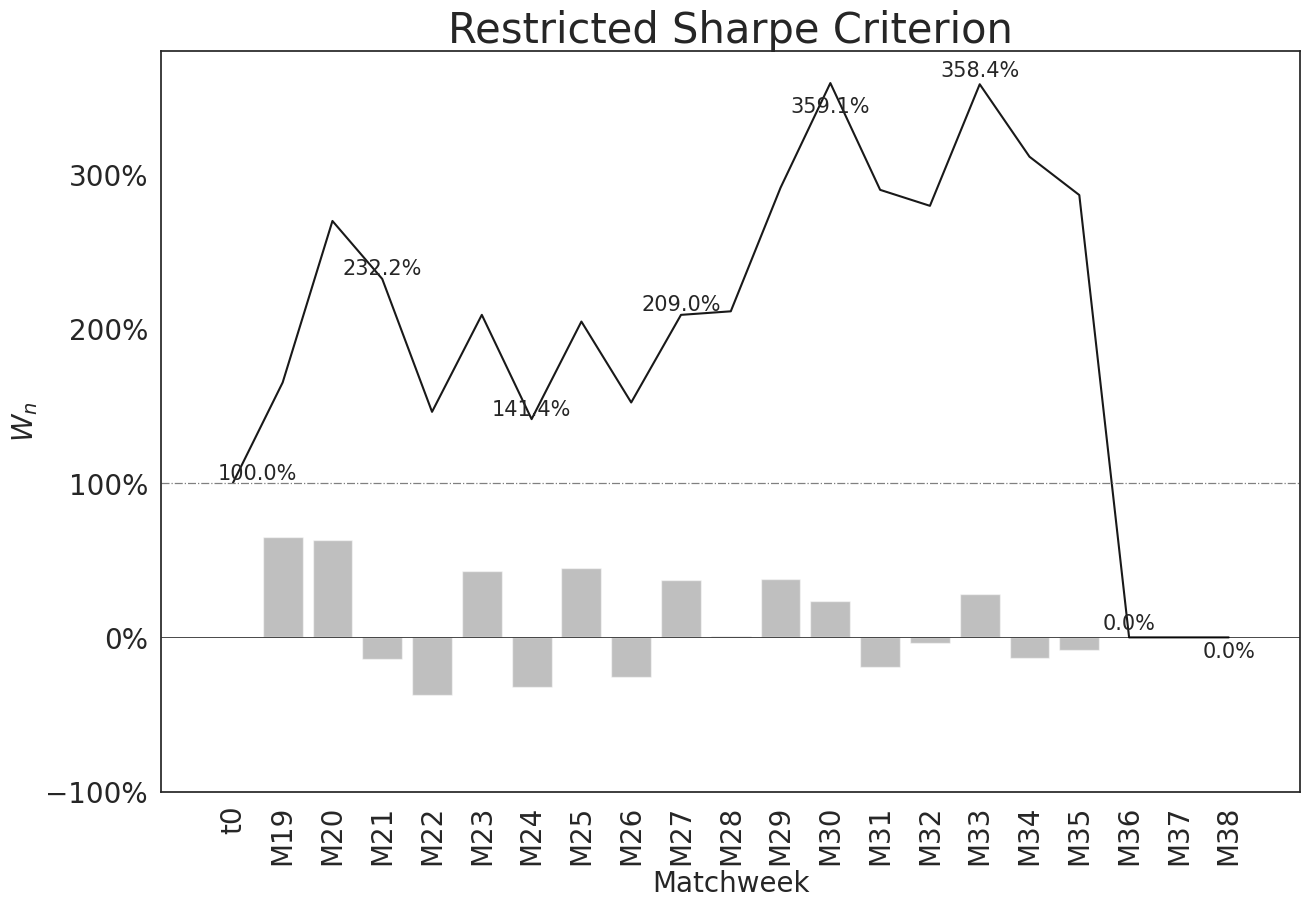}
  \label{fig:3.2.3}
\end{subfigure} %
\begin{subfigure}{.52\textwidth}
  \centering
  \includegraphics[width=0.9\linewidth]{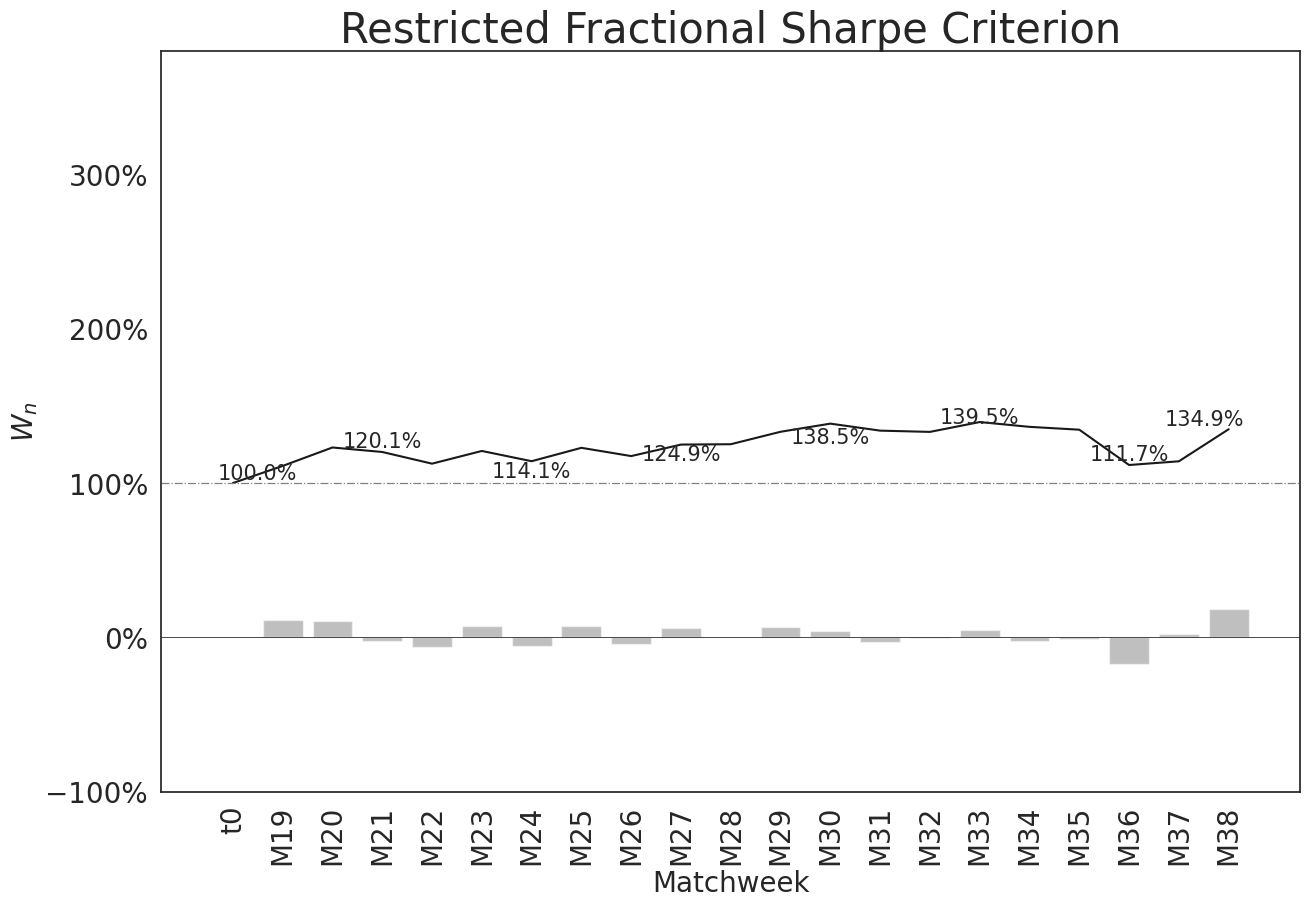}
  \label{fig:3.2.4}
\end{subfigure} %
\caption{Plots depicting the wealth $W_{n}$ under restricted strategies for both quadratic and logarithmic utilities.}
\label{fig:restricted_models}
\end{figure}

The table below presents the portfolios derived from the Kelly Criterion and the Sharpe Ratio Criterion for the final matchweek of the English Premier League. It is noteworthy that the Kelly criterion never wagers the entire wealth, unlike the Sharpe Ratio criterion. Additionally, while the two criteria allocate similar amounts for each event, the magnitude of the largest bet differs between them.

\begin{table}[ht]
\captionsetup{font=footnotesize}
\centering
\resizebox{16cm}{!}{
\begin{tabular}{lcccccccccc} 
\hline
\textbf{Home}                                            & Arsenal    & Aston Villa & Fulham     & Leeds      & Leicester  & Liverpool  & Man. City   & Sheffield  & Westham     & Wolves      \\
\textbf{Away}                                           & Brighton   & Chelsea     & Newcastle  & West Brom  & Tottenham  & Crystal    & Everton    & Burnley    & Southampton & Man. United  \\ 
\hline
\textbf{$\mathscr{e}$}                          & \textbf{D} & \textbf{H}  & \textbf{A} & \textbf{A} & \textbf{A} & \textbf{A} & \textbf{H} & \textbf{A} & \textbf{A}  & \textbf{A}  \\
\textbf{$\ou$ }                            & 4.46       & 7           & 3.41       & 7          & 3.5        & 18.32      & 1.47       & 2.45       & 5.03        & 2.75        \\
\textbf{$\hat{p}$ }                             & 26.25\%    & 31.66\%     & 41.48\%    & 26.85\%    & 31.58\%    & 10.17\%    & 75.39\%    & 48.46\%    & 25.11\%     & 41.88\%     \\
\textbf{$\mathbb{E}[\varrho]$} & 17.07\%    & 121.65\%    & 41.45\%    & 87.95\%    & 10.53\%    & 86.24\%    & 10.82\%    & 18.73\%    & 26.31\%     & 15.17\%     \\
\multicolumn{11}{c}{\textbf{Strategies}}                                                                                                                                                      \\
\textbf{Kelly}                                            & 3.23\%     & 17.89\%     & 12.77\%    & 12.05\%    & 2.65\%     & 3.93\%     & 14.33\%    & 8.56\%     & 4.47\%      & 5.61\%      \\
\textbf{Sharpe}                                        & 4.44\%     & 11.50\%     & 14.72\%    & 9.16\%     & 3.99\%     & 2.82\%     & 27.05\%    & 12.52\%    & 5.54\%      & 8.26\%      \\
\hline
\end{tabular}
}
\caption{\label{tab:table_mtchw_kelly_mktz} Restricted strategies for the final day of the 20/21 EPL. The events $e$ denote bets on home win \textbf{H}, draw \textbf{D} or away win \textbf{A}.}
\end{table}

\subsubsection{Portfolios' Performance}
The metrics for the eight portfolios for the 20 fixtures are summarized below, including the metrics \textit{pval bets} and \textit{pval wealth}. The \textit{pval bets} metric represents the p-value of the hypothesis test that the average of the betting outcomes is not positive. Similarly, the \textit{pval wealth} metric is the p-value for the hypothesis that the wealth obtained in the matchweeks is not positive.

\begin{table}[ht]
\captionsetup{font=footnotesize}
\centering
\resizebox{14cm}{!}{
\begin{tabular}{llcccccccc} 
\toprule
\multirow{3}{*}{\textbf{Model}}    & \textbf{Strategy}              & \textbf{Sharpe} & \textbf{Kelly} & \textbf{Kelly} & \textbf{Sharpe} & \textbf{Kelly} & \textbf{Sharpe} & \textbf{Kelly} & \textbf{Sharpe}  \\
                                    & \textbf{Restricted}             & Yes                & Yes            & No            & No                & Yes            & No                & No            & Yes                 \\
                                    & \textbf{Fracción}                & 17\%               & 17\%           & 17\%           & 17\%               & 100\%          & 100\%              & 100\%          & 100\%               \\ 
\midrule
\multirow{10}{*}{\textbf{Metrics}} & \textbf{Final Wealth}                      & \textbf{135\%}     & \textbf{111\%} & \textbf{102\%} & \textbf{97\%}      & \textbf{24\%}  & \textbf{16\%}      & \textbf{3\%}   & \textbf{0\%}        \\
                                    & \textbf{Number of Bets}           & 196                & 195            & 319            & 323                & 195            & 323                & 319            & 176                 \\
                                    & \textbf{Average Stake} & 17\%               & 14\%           & 17\%           & 17\%               & 83\%           & 100\%              & 98\%           & 100\%               \\
                                    & \textbf{Hits}        & 33\%               & 33\%           & 30\%           & 30\%               & 33\%           & 30\%               & 30\%           & 32\%                \\
                                    & \textbf{Sharpe Average}                  & 0.62               & 0.58           & 0.64           & 1.41               & 0.58           & 1.41               & 0.64           & 0.62                \\
                                    & \textbf{Log-Growth Average}                      & 0.06               & 0.06           & 0.07           & 0.04               & 0.21           & 0.16               & 0.23           & -$\infty$                \\
                                    & \textbf{Volatility Average}                & 0.01               & 0.02           & 0.02           & 0.00               & 0.56           & 0.13               & 0.54           & 0.38                \\
                                    & \textbf{Betting pval}           & 0.21               & 0.35           & 0.42           & 0.50               & 0.35           & 0.50               & 0.42           & 0.35                \\
                                    & \textbf{Wealth pval}            & 0.16               & 0.33           & 0.41           & 0.50               & 0.33           & 0.50               & 0.41           & 0.31                \\
\bottomrule
\end{tabular}
}
\caption{\label{tab:table_comparative_finance} Summary statistics of the performance for each of the portfolios throughout the test period.}
\end{table}
\section{Conclusions}

This study aimed to address several key aspects\nospacecolon \ first, identifying the optimal betting strategy for a rational gambler seeking to maximize expected utility, taking into account logarithmic or quadratic utilities. Second, analyzing the characteristics and performance of complete and restricted betting strategies to gain insights into their respective qualities. Third, exploring the qualitative and quantitative differences between portfolios constructed based on the Kelly Criterion and the Sharpe Ratio Criterion in a real-life betting scenario. Fourth, developing statistical learning models to forecast outcomes in the English Premier League. Notably, this study made significant contributions by introducing a matrix-based formulation (\ref{eq2:multiple_kelly}) for the multivariate Kelly Criterion and formulating the optimization problem for the multiple and simultaneous Kelly Criterion (\ref{eq2:multiple_simult_kelly}), enhancing the understanding and applicability of these approaches in practical settings.

The successful demonstration of the first objective of this study involved the development of a systematic method to identify optimal bets within the framework of $(D, E, C, (\succeq))$. The method followed a step-by-step approach\nospacecolon \ first, by defining the set $E$ to determine the elements of $C$, and subsequently constructing the set $D$ from these two sets. Second, the probabilities of the joint events were estimated. Third, the optimal decision $d_{\ell_{*}}$ was determined by maximizing the expected utility of returns, denoted as $\EUT{R(\underline{\ell})}$, while ensuring that $d_{\ell_{*}}$ belongs to the set $D$. Finally, appropriate metrics were defined to assess the performance of the strategies in terms of returns.

Regarding the second aspect investigated, it was observed that although the restricted strategies exhibited better convergence and higher returns compared to the complete strategies, they also displayed higher variance and lower diversification. Notably, in extreme cases, the Sharpe Ratio criterion led to the possibility of the player's ruin, as evidenced by the complete loss of funds two weeks prior to the conclusion of the study. Moreover, it was theoretically established that the advantages of exploiting arbitrage opportunities diminish when stakes are limited. Furthermore, the narrowing of the set of possible bets in restricted strategies resulted in a compromise of the fundamental properties of the criteria. Both the Kelly Criterion and the Sharpe Ratio Criterion exhibit similarities and differences in their portfolio outcomes. Firstly, unlike portfolios based on the Sharpe Ratio, Kelly-based portfolios offer protection against the risk of player ruin. For instance, on average, the full investment strategy and the Kelly-constrained strategy wagered 98\% and 83\% of the total wealth, respectively. Secondly, it is noteworthy that the logarithmic growth achieved through the Kelly Criterion is not invariant to fractional bets, which sets it apart from the Sharpe Ratio approach. Consequently, fractionalizing the Kelly strategy prior to optimization results in a distinct strategy compared to the fractional strategy $f\underline{\ell}$ derived post-optimization.

Thirdly, it is noteworthy that portfolios utilizing logarithmic utilities tend to exhibit relatively lower levels of diversification compared to portfolios employing quadratic utilities, as a proportion of the total stake invested in the portfolio. This observation was addressed by examining the maximum stake per matchweek relative to the total amount wagered in the portfolio, revealing that the Kelly Criterion maximum bet averaged 29.7\% compared to the maximum Sharpe Ratio's bet average of 25.9\%. Consequently, logarithmic portfolios tend to display higher volatility, resulting in more pronounced gains or losses. However, when considering the total budget of \$1, the maximum Kelly Criterion stakes average 1.9\% less than the Sharpe Ratio approach. This discrepancy arises due to the logarithmic nature of the growth function $G(\underline{\ell})$, which tends towards negative infinity as $\sum_{i}\ell_{i}$ approaches 1, as demonstrated by Thorp \cite{thorp1969optimal}.

Fourth, while the constituent elements of the portfolios are similar for both the Kelly Criterion and the Sharpe Ratio Criterion (see Table \ref{tab:table_mtchw_kelly_mktz}), the strategies themselves are not proportionally aligned between the two methods. This divergence is evident even in extreme cases such as arbitrage, where notable differences are observed in the resulting portfolios (see Table \ref{tab:table_arbitrage}).

In regards to the deep learning model, several considerations need to be addressed. Despite the advantages in terms of predictive power, flexibility, and relaxed assumptions offered by neural networks in modeling sporting events, it is important to acknowledge that the present study faced limitations due to the low volume of available data. Nevertheless, employing a deep learning approach still represents an improvement over traditional multinomial regression for sports modeling. Furthermore, the selection of variables used in the model remains an area of ongoing investigation. It is evident that there is a deficiency in the number of variables available to accurately determine match outcomes, particularly at the individual player level. Despite this limitation, Zimmerman suggests that there exists an empirical benchmark of 75\% in terms of predictive accuracy for sporting events \cite{zimmermann2013predicting}. Moreover, Hub{\'a}{\v{c}}ek emphasizes that the selection of variables holds greater significance than the specific statistical model employed used \cite{hubavcek2019exploiting}. These perspectives underscore the importance of further refining the variable selection process to enhance the predictive capabilities of the model.

\begin{figure}[ht]
\begin{subfigure}{.52\textwidth}
  \centering
  \includegraphics[width=0.8\linewidth]{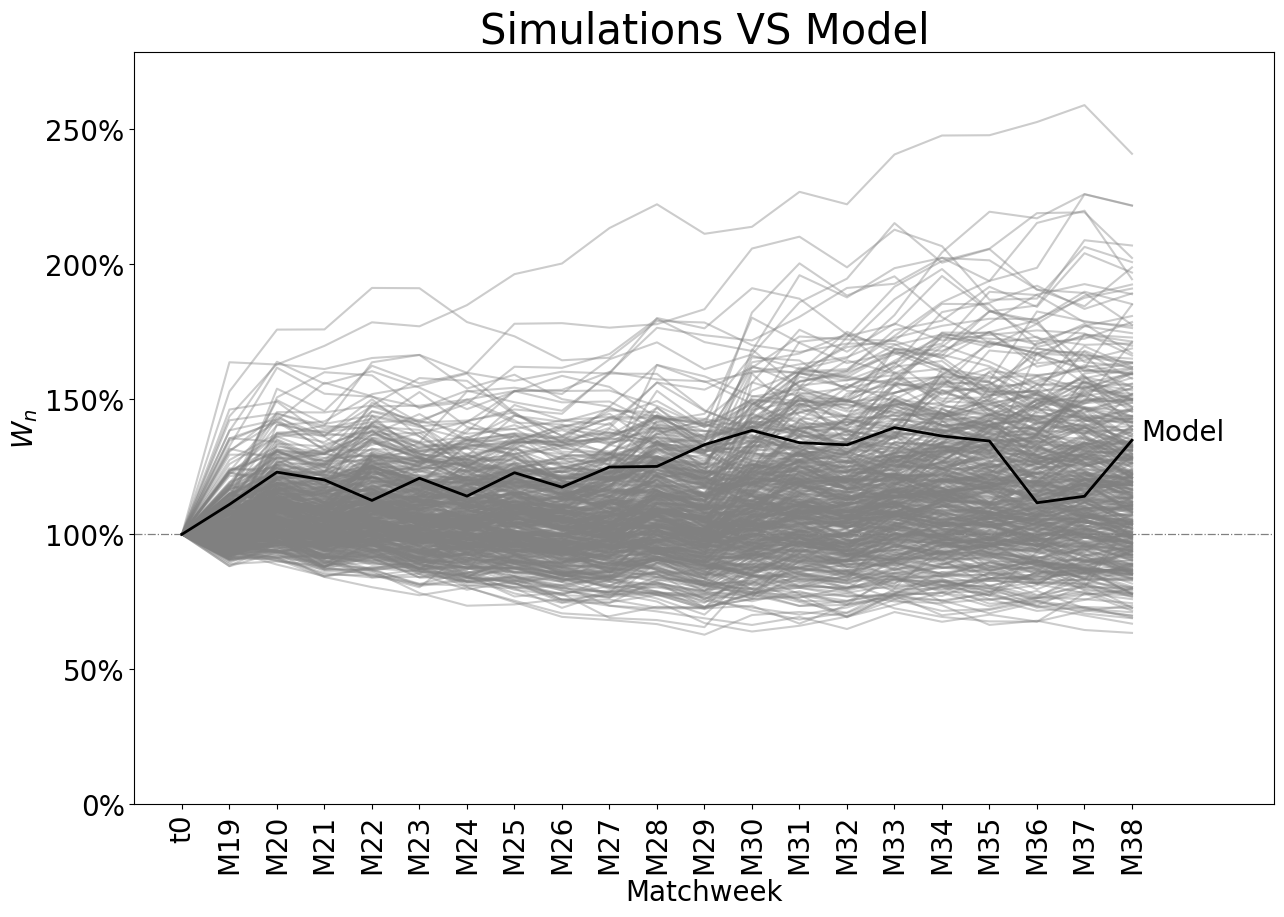}
  \label{fig:7.1splt1}
\end{subfigure}%
\begin{subfigure}{.52\textwidth}
  \centering
  \includegraphics[width=0.8\linewidth]{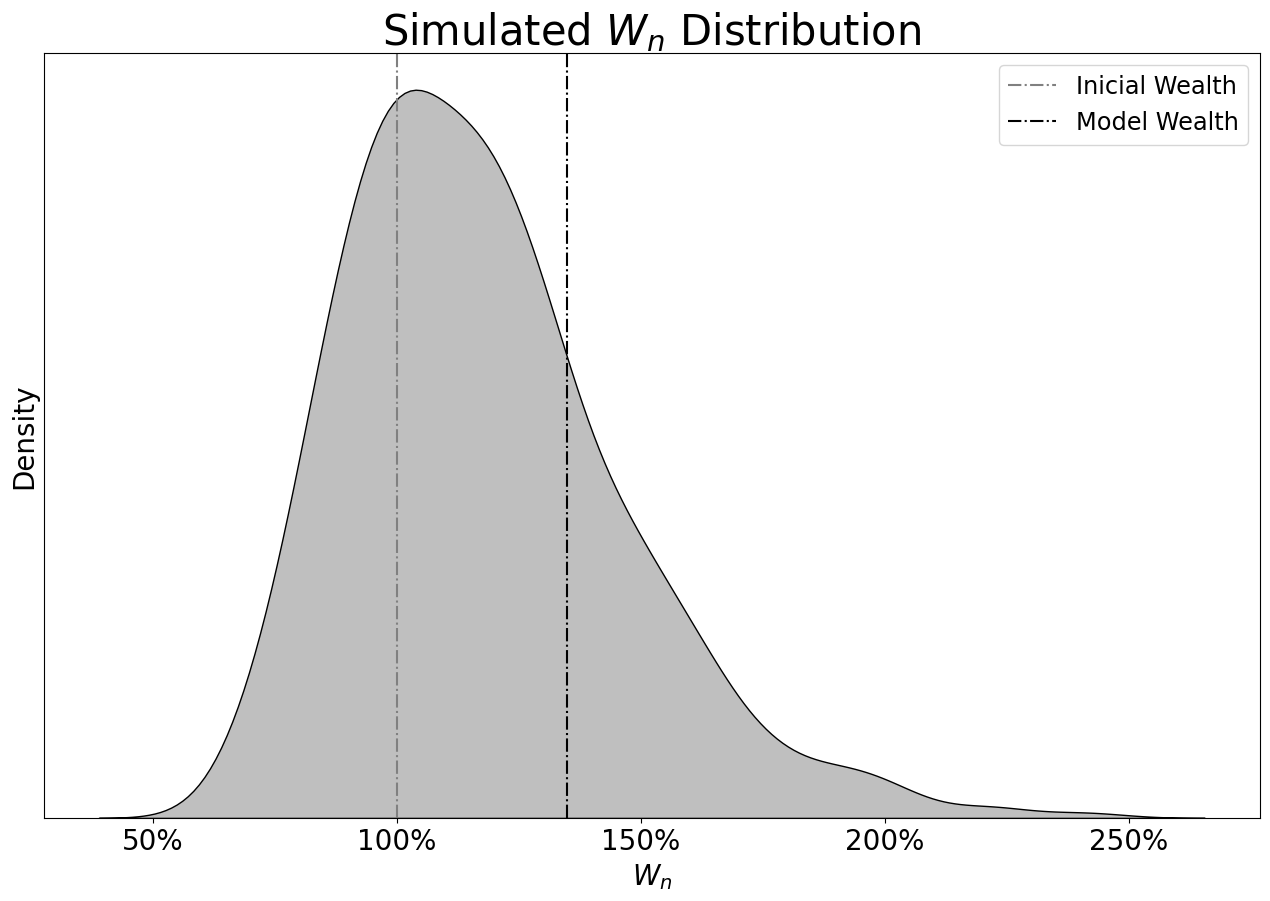}
  \label{fig:7.1splt2}
\end{subfigure} %
\caption{Inter-temporal and final distribution of 500 simulated strategies. The final wealth of these strategies is compared with that of the best model.}
\label{fig:monkeys_vs_model}
\end{figure}

Finally, it is imperative to validate the underlying assumptions of the models. The cornerstone of the developed method relies on the premise that the actual probabilities of events are known. However, upon examining the predictions' variability in Table \ref{tab:table_models_metrics} and the classification errors illustrated in Table \ref{tab:mat_conf}, doubts arise regarding the veracity of this assumption. Notably, the model's a priori cross entropy was anticipated to be 1.0318, whereas the empirical entropy based on frequencies amounts to 1.0631. Furthermore, the maximum entropy for a three-event phenomenon is calculated to be 1.0986. Consequently, the certainty surrounding this assumption is called into question.

Furthermore, it is crucial to evaluate whether the model's performance of 35.8\% can be attributed to "luck" or "skill." To address this, the model is subjected to a test against 500 Dirichlet simulations with unit parameters, conducted within the same test period but without restrictions and employing strategies bounded to the same fraction $f$ as the model in question. The results reveal that, on average, the model outperforms 78 out of 100 simulations. This serves as compelling evidence that the model's performance extends beyond mere chance. Nevertheless, despite investigating the relationship between log-growth and performance for the simulations, no linear evidence supporting such a connection is found (p-value of 0.595).


\subsection{Future Research}
The findings and hypotheses explored in this study open up various avenues for future research. These avenues span both financial methodologies and predictive modeling in the context of sports betting.

From a financial perspective, it would be of great theoretical and empirical interest to examine the long-term behavior of returns when encountering arbitrage opportunities. Exploring regularization techniques, particularly in the context of $L_{2}$ regularization for betting portfolios, could also yield valuable insights. Additionally, investigating the divergences in portfolio composition and returns between restricted and full strategies, considering both quadratic and logarithmic utilities, through simulation studies would provide further understanding.

In the realm of predictive modeling, there are multiple aspects worth exploring. One avenue is adopting a Bayesian perspective to analyze the prediction problem, complementing the frequentist approach employed in this research. Furthermore, incorporating player-level data in addition to aggregate team data could enhance the predictive model's accuracy and granularity. The inclusion of data from other soccer leagues could also contribute to a more comprehensive and robust modeling approach.

It is important to acknowledge that the work presented in this study is grounded in the Von Neumann and Morgenstein's Axioms of Preference. However, it is well-known that these axioms may not always hold in reality. Exploring the disparities between theoretical preferences and empirical observations, given the same available information, would shed light on the limitations and challenges associated with relying solely on axiomatic models. Furthermore, investigating optimal policies under Reinforcement Learning models could provide valuable insights into the dynamic decision-making processes within the realm of sports betting.

Overall, the future research directions outlined above have the potential to further advance the understanding of financial strategies, predictive modeling approaches, and the complex dynamics of decision-making in sports betting beyond this paper.
\section*{Appendix}
The apendix provide a list and description of the three databases utilized in this study for the estimation of probabilities and obtaining odds from various bookmakers. The variables employed in the prediction of match outcomes, with a dagger ($\dagger$) indicating their usage, and the target variable marked by an asterisk ($*$), are also specified.

\begin{multicols}{2}

\subsection*{Football Data U.K.}

The "Football Data U.K." database contains records representing individual matches played in the English Premier League from the 2013/2014 season through the 2020/2021 season. The database includes a range of variables that provide valuable insights into match characteristics and team performance. However, it should be noted that a subset of variables was excluded from the analysis for matches prior to 2019, as these variables were not available in the reported files during that time period.
\subsubsection*{Fetched Variables}

\begin{itemize}[leftmargin=*]
    \setlength{\itemindent}{0em}
    
    \item date\nospacecolon \  Date in day/month/year on which the match took place.
    \item hometeam\nospacecolon \  Home team name.
    \item awayteam\nospacecolon \  Name of the team playing as visitor.
    \item fthg\nospacecolon \  Goals scored at the end of the home team's game.
    \item ftag\nospacecolon \  Goals scored at the end of the game by the away team.
    \item ftr\nospacecolon \  Final score. The possible values are\nospacecolon \  \textit{H, D, A}. Which represent that the home team won, drew or that the visiting team won, respectively.
    \item referee\nospacecolon \  Name of the main referee who directed the match.
    \item \_h, \_d, \_a\nospacecolon \  The bookmakers' odds for the possible outcomes of the match. The information is collected on Tuesdays and Fridays. The bookmakers are:
        \begin{itemize}
            \item b365\nospacecolon \  Bet365.
            \item bw\nospacecolon \  Bwin.
            \item iw\nospacecolon \  Interwetten.
            \item ps\nospacecolon \  Pinnacle Sports.
            \item vc\nospacecolon \  Victor Chandler.
            \item wh\nospacecolon \  William Hill.
        \end{itemize}
    \item$\dagger$psch, pscd, psca\nospacecolon \  Pinnacle Sports' final odds, that is, an instant before the match starts, for the result that the home team wins, draws or that the away team wins.

\end{itemize}

\subsubsection*{Generated Variables}

\begin{itemize}[leftmargin=*]
    \setlength{\itemindent}{0em}
    
    \item$\dagger$matchweek\nospacecolon \  The day on which the respective matches are played.
    \item$*$result\nospacecolon \  The variable \textit{ftr} renamed.
    \item season\nospacecolon \  Season in which they are playing. If the season is 2013/2014, 13 is captured. 
    \item maxo\_\nospacecolon \  The maximum odds for the three possible outcomes (H, T, V) of the six bookmakers mentioned at the beginning. \textit{Note}\nospacecolon \  Pinnacle Sports final odds are not considered. \textit{Note 2}\nospacecolon \  These are the odds used for betting.
    \item market\_tracktake\nospacecolon \  The market commission. That is, the sum of the inverse of the maximum odds found for each outcome.
    \item diff\_\nospacecolon \  The relative difference between the odds -without the respective commission of the collected and the final Pinnacle Sports odds.
\end{itemize}

\subsection*{Understats}

The Understats database provides match-level data for each team in the English Premier League. Each record in the table represents a match that a team has played, rather than the match itself. This means that if there are 380 matches in the Premier League in a season, there will be 760 records in this table for that season.

It is important to note that observations for teams on the first matchweek of each season were removed from the dataset. This is because the variance between the last game of the previous season and the first game of the current season tends to be very large due to changes in the player market and initial team performances, as mentioned on the present work. Removing these observations helps to avoid potential biases in the data caused by these factors.

\subsubsection*{Fetched Variables}

\begin{itemize}[leftmargin=*]
    \item h\_a\nospacecolon \  Character that represents whether the team plays as home or away team, whose values are \textit{a}, \textit{h}, respectively.
    \item xG\nospacecolon \  Number of goals expected in the match by the team. 
    \item xGA\nospacecolon \  Number of goals expected in the match by the opposing team.
    \item npxG\nospacecolon \  Number of expected goals in the match by the team without taking penalties into account.
    \item npxGA\nospacecolon \  Number of goals expected in the match by the opposing team without taking penalties into account.
    \item npxGD\nospacecolon \  Difference between npxG and npxGA.
    \item deep\footnote{These statistics present great inconsistencies with respect to the official Understats page. Likewise, since they were obtained through an R and Python package, the methodology with which these variables were obtained is unknown.}\nospacecolon \  Number of passes completed by the team in the last quarter of the court -on the opposing team's side.
    \item deep\_allowed\footnotemark[1]\nospacecolon \  Number of passes completed by the opposing team in the last quarter of the court -on the team's side-.
    \item scored\nospacecolon \  Number of goals scored by the team in the match.
    \item missed\nospacecolon \  Number of goals conceded by the opposing team in the match.
    \item xpts\nospacecolon \  Number of expected points. It is the expected result for the team.
    \item result\nospacecolon \  Result of the match for the team. Possible values are \textit{w, d, l} representing that the team won, drew or lost the match, respectively.
    \item date\nospacecolon \  Date in year-month-day when the match took place.
    \item wins, draws, loses\nospacecolon \  Dummy variables representing whether the team won, drew or lost the match, respectively.
    \item pts\nospacecolon \  Points obtained by the result of the match for the team. Winning, drawing or losing awards 3, 1 and 0 points, respectively.
    \item ppda.att\footnotemark[1]\nospacecolon \  Total passes made by the team when attacking divided by the number of defensive actions of the opposing team (interceptions + tackles + fouls). Metric suggested by Colin Trainor.
    \item ppda.def\footnotemark[1]\nospacecolon \  Total passes made by the team when defending divided by the number of defensive actions by the opposing team.
    \item ppda.att\footnotemark[1]\nospacecolon \  Total passes made by the opposing team when attacking divided by the number of the team's defensive actions (interceptions + tackles + fouls). Metric suggested by Colin Trainor.
    \item ppda\_allowed.def\footnotemark[1]\nospacecolon \  Total passes completed by the opposing team while defending divided by the number of defensive team actions.
    \item team\_id\nospacecolon \  Id with which Understats identifies the team.
    \item team\_name\nospacecolon \  Name with which Understats identifies the team.
    \item league\_name\nospacecolon \  Name by which Understats identifies the league.
    \item year\nospacecolon \  Season number of the match. If the season is 2013/2014, 2013 is captured.
    \item matchweek\nospacecolon \  Day of the season of the current match. There are 38 matchweeks in total.
\end{itemize}

\subsubsection*{Generated Variables}
\begin{itemize}[leftmargin=*]
    \item$\dagger$position\_table:The position in tables for the current day before the games.
    \item$\dagger$total\_points\nospacecolon \  The total points of the team for the current day before the games.
    \item $\dagger$promoted\_team\nospacecolon \  Dummy variable indicating whether the team was promoted to the EPL in the current season.
    \item$\dagger$big\_six\nospacecolon \  Dummy variable that indicates whether the team is a Big Six. That is, if the team is Arsenal, Chelsea, Liverpool, Manchester City, Manchester United or Tottenham.
    \item season\nospacecolon \  The last two digits of the variable \texttt{year}.
    \item$\dagger$npxGD\_ma\nospacecolon \  It is the weighted average, with $\xi = 0.1$, of the \textit{npxGD} from the first record of the team until the current matchday prior to the games.
  \item$\dagger$npxGD\_var\nospacecolon \  Is the weighted variance, with $\xi = 0.1$, of the \textit{npxGD} from the team's first record to the present day prior to the games.    
    
\end{itemize}

\subsection*{SoFIFA}

The author utilized web scraping techniques to download tables from the SoFIFA page for the English Premier League. The data obtained represents the statistics of teams for each week of each season, with a one-matchday delay to reflect the team's status in the corresponding week prior to the matches to be played.

In the dataset, each record represents a team in a specific week. However, there are cases where more than one team's statistics were reported in a single week. To address this, only the last record reported on the page for each week was considered. In instances where there is no record available for a particular week, data from the previous week was utilized instead.

\subsubsection*{Fetched Variables}
\begin{itemize}[leftmargin=*]
    \item name\_team\nospacecolon \  Name under which SoFIFA identifies the team.
    \item id\nospacecolon \  Id with which SoFIFA identifies the team.
    \item$\dagger$ova\nospacecolon \  Rating from 1 to 100 of the team's overall performance up to that week.
   \item$\dagger$att\nospacecolon \  Rating from 1 to 100 of the team's attack up to that week.
    \item$\dagger$mid\nospacecolon \   Rating from 1 to 100 of the team's average up to that week.
    \item$\dagger$def\nospacecolon \  Rating from 1 to 100 of the team's defense up to that week.
    \item$\dagger$transfer\_budget\nospacecolon \  Budget for the transfer market, in millions of euros, of the team for that season.
    \item speed\footnote{Despite being variables with excellent information, for the 2019 seasons onwards, SoFIFA stopped updating these values and they became constant for all teams. So these variables were no longer, in their entirety, informative.}\nospacecolon \  Type of speed the team plays with.
    \item dribbling\nospacecolon \  Type of the number of dribbles with which the team plays\footnotemark[2]..
    \item passing\nospacecolon \  Type of passes with which the team plays. They can be very risky, normal or safe passes.\footnotemark[2].
    \item positioning\nospacecolon \  Formation with which the team plays.
    \item crossing\footnotemark [2]\nospacecolon \  Type of band changes in the passes with which the team plays.
    \item aggression\footnotemark[2]\nospacecolon \  Aggressiveness with which the team defends.
    \item pressure\footnotemark[2]\nospacecolon \  Pressure with which the team defends.
    \item team\_width\nospacecolon \  Width of the formation with which the team plays.
    \item defender\_line\footnotemark[2]\nospacecolon \  Type of mark with which the team defends.
    \item dp\nospacecolon \  Number of the team's domestic prestige. It is rated from 1 to 20.
   \item$\dagger$ip\nospacecolon \  Number of the international prestige of the team. It is graded from 1 to 20.
    \item players\nospacecolon \  Number of players registered by the team to play in the current EPL season.
    \item$\dagger$saa\nospacecolon \  Average age of the starting roster for that season as of that date.
    \item taa\nospacecolon \  Average age of the team for that season at that date.
    \item date\nospacecolon \  Date in year-month-day when SoFIFA published the EA Sports data of the teams.
    \item fifa\nospacecolon \  Name and number of the EA Sports FIFA video game.
    \item year\_week\nospacecolon \  Date in year-week when SoFIFA published the EA Sports data of the teams. The date is in ISO 8601 format.

\end{itemize}

\subsubsection*{Generated Variables}
No new variables were transformed or generated.

\end{multicols}

\subsection*{Main Database}
The main database used in the machine learning models consists of various variables, each undergoing a specific transformation before being used to train the neural networks. It should be noted that the standardization or normalization transformations applied to the variables have a time window of one matchweek, meaning that the calculations are based on data from the same week. These transformations are important for ensuring that the variables are appropriately scaled and prepared for input into the neural networks.
\begin{multicols}{3}
\begin{enumerate}[leftmargin=*]
    \item matchweek\footnote{Normalized. That is for say $s = (x - x_{(1)}) / (x_{(n)} - x_{(1)})$.}

    \item position\_table\_home\footnote{Inversely normalized. In other words $s = (x_{(n)} - x) / (x_{(n)} - x_{(1)})$.}
    \item total\_pts\_\footnote{Standarized. i.e. $t = (x - \bar{x}) / \hat{s}$.}
    \item npxGD\_ma\_home
    \item npxGD\_var\_home
    \item big\_six\_home
    \item promoted\_team\_home

    \item position\_table\_away\footnotemark[15]
    \item total\_pts\_away\footnotemark[16]
    \item npxGD\_ma\_away
    \item npxGD\_var\_away
    \item big\_six\_away
    \item promoted\_team\_away

    \item ova\_home\footnotemark[16]
    \item att\_home\footnotemark[16]
    \item mid\_home\footnotemark[16]
    \item def\_home\footnotemark[16]
    \item transfer\_budget\_home\footnote{Normalized, with the maximum observation cliffed at 100.}
    \item ip\_home\footnotemark[16]
    \item saa\_home\footnotemark[14]

    \item ova\_away\footnotemark[16]
    \item att\_away\footnotemark[16]
    \item mid\_away\footnotemark[16]
    \item def\_away\footnotemark[16]
    \item transfer\_budget\_away\footnotemark[17]
    \item ip\_away\footnotemark[16]
    \item saa\_away\footnotemark[14]

    \item proba\_h
    \item proba\_d
    \item proba\_a
\end{enumerate}
\end{multicols}

\bibliographystyle{unsrt}  
\bibliography{references.bib}

\end{document}